\documentclass[lettersize,journal]{IEEEtran}

\usepackage{amsmath,amsfonts}
\usepackage{algorithmic}
\usepackage{algorithm}
\usepackage{array}

\usepackage[dvipsnames]{xcolor}
\usepackage{textcomp}
\usepackage{stfloats}
\usepackage{siunitx}
\usepackage{url}
\usepackage{verbatim}
\usepackage{graphicx}
\usepackage{rotating}
\usepackage[export]{adjustbox}
\usepackage{cite}
\usepackage{multirow}
\usepackage{pdflscape}
\usepackage{pdfpages}
\usepackage{subfig}
\usepackage[utf8]{inputenc}
\usepackage[T1]{fontenc}
\begin{document}

\title{Sneak Path Current Modeling in Memristor Crossbar Arrays for Analog In-Memory Computing}

%title{Parametric Closed Form Model for Accurate Sneak Path Prediction in Memristor Crossbar Arrays}
%\thanks{* Equal contribution.}
%\thanks{This work was supported in part by the National Science Foundation under grants 2510192, 2502054, 2428981, 2420994, 2427766, 2408064, 2247343, and 2218046. This material was also based upon work supported by the U.S. Department of Energy, Office of Science, Office of Advanced Scientific Computing Research (ASCR), under Award Number(s) DE-SC0025561.}}

\author{Shah Zayed Riam*, Zhenlin Pei*,~\IEEEmembership{Graduate Student Member,~IEEE,} Kyle~Mooney,~\IEEEmembership{Graduate Student~Member,~IEEE}, Chenyun Pan,~\IEEEmembership{Senior Member,~IEEE,} Na~Gong,~\IEEEmembership{Senior Member,~IEEE,} and~Jinhui~Wang,~\IEEEmembership{Senior Member,~IEEE.}

%\author{\IEEEauthorblockN{Shah Zayed Riam*, Zhenlin Pei*, Kyle Mooney, Chenyun Pan, Na Gong, and Jinhui Wang}

%\IEEEauthorblockA{\textit{Electrical and Computer Engineering, The University of Alabama, Tuscaloosa, AL, USA}\\
%Tuscaloosa, AL, USA} \\
%\textit{The University of Alabama}\\
%Tuscaloosa, AL, USA \\
%jwang231@ua.edu
%}
%\and
%\IEEEauthorblockN{Zhenlin Pei}
%\IEEEauthorblockA{\textit{Electrical and Computer Engineering} \\
%\textit{The University of Alabama}\\
%Tuscaloosa, AL, USA \\
%}
%\and
%\IEEEauthorblockN{Jinhui Wang}
%\IEEEauthorblockA{\textit{Electrical and Computer Engineering} \\
%\textit{The University of Alabama}\\
%Tuscaloosa, AL, USA \\
%}
\thanks{Manuscript received November 2025. This work is supported in part by the National Science Foundation under grants 2510192, 2502054, 2428981, 2420994, 2427766, 2408064, 2247343, and 2218046.  This material was also based upon work supported by the U.S. Department of Energy, Office of Science, Office of Advanced Scientific Computing Research (ASCR), under Award Number(s) DE-SC0025561.
\textit{(Corresponding author: Jinhui Wang)}}

\thanks{Shah Zayed Riam, Zhenlin Pei, Kyle Mooney, Na Gong, and Jinhui Wang are with the Department of Electrical and Computer Engineering, The University of Alabama, Tuscaloosa, AL 35487 USA (e-mail: jwang231@ua.edu).}

\thanks{Chenyun Pan is with the Department of Electrical Engineering, The University of Texas at Arlington, Arlington, TX 76010 USA.}

\thanks{*Denotes equal contribution and co-first author.}

\thanks{Color versions of one or more of the figures in this article are available online at http://ieeexplore.ieee.org}}

\markboth{IEEE Journal on Emerging and Selected Topics in Circuits and Systems,~Vol.~XX, No.~XX, XX~2025}%
{Shell \MakeLowercase{\textit{et al.}}: A Sample Article Using IEEEtran.cls for IEEE Journals}

\maketitle

\begin{abstract}
 Memristor crossbar arrays have emerged as a key component for next-generation non-volatile memories, artificial neural networks, and analog in-memory computing (IMC) systems. By minimizing data transfer between the processor and memory, they offer substantial energy savings. However, a major design challenge in memristor crossbar arrays is the presence of sneak path currents, which degrade electrical performance, reduce noise margins, and limit reliable operations. This work presents a closed-form analytical framework based on 1.4nm technology for accurately estimating sneak path currents in memristor crossbar arrays. The proposed model captures the interdependence of key design parameters in memristor crossbar arrays, including array size, ON/OFF ratio of memristors, read voltage, and interconnect conditions, through mathematically derived relationships. It supports various practical configurations, such as different data patterns and connection strategies, enabling rapid and comprehensive sneak path current modeling. The sensitivity analysis includes how design parameters influence sneak path current and noise margin loss, underscoring the trade-offs involved in scaling crossbar arrays. Validation through SPICE simulations shows that the model achieves an error of less than 10.9\%\ while being up to 4784 times faster than full circuit simulations. This analytical framework offers a powerful tool for quantitative assessment and pre-design/real-time optimization of memristor-based analog in-memory computing (IMC) architectures.

\end{abstract}


\section{Introduction}
\label{sec:Introduction}
\IEEEPARstart{M}EMRISTOR, widely recognized as the fourth fundamental circuit element, has become central to the development of next-generation non-volatile memory technologies and emerging analog in-memory computing (IMC)
systems, addressing the scaling and efficiency challenges of traditional CMOS-based memories~\cite{chua2003memristor,pourbakhsh2016sizing,gong2012clock,edstrom2019data,gong2014variation,chen2018viewer,khan2023pawn}. The primary architectural manifestation of memristor-based memory is the high-density passive crossbar array, where each cell is located at the intersection of perpendicular wordlines and bitlines \cite{vourkas2015memristive,shin2011analysis}. This structure provides exceptional storage density since each cell can be scaled to the nanometer regime without requiring a transistor~\cite{kim2012functional,jo2009high}, thereby offering higher bit-per-area ratios than conventional memories~\cite{lastras2018resistive}. Combined with its non-volatility and fast read capability, the memristor crossbar helps bridge the performance gap in modern memory hierarchies~\cite{uppaluru2025carbon,6730965,oli2022stuck,oli2022reliability}. Additionally, memristor plays a crucial role in analog computing by enabling fast and direct hardware implementation of vector matrix multiplication in-memory, such as in a neuromorphic computing system. Memristors can perform computation intrinsically through Ohm's and Kirchhoff's laws when placed in crossbar arrays. The output current produced is in the form of weighted summations of input voltage and stored conductance of the memristor. This characteristic of analog IMC results in highly parallel, energy-efficient acceleration for signal processing and Artificial Intelligence (AI) tasks, as it removes the need for frequent data transfer between memory and processors \cite{hu2018memristor}.

However, a key challenge in fully realizing the advantages of passive memristor crossbar arrays is the sneak path current~\cite{vontobel2009writing}. Sneak paths are unwanted parallel current routes that occur during read operations, diverting current away from the target cell. In unoptimized crossbars, sneak path currents can dominate the total current by exceeding 50\%, leading to substantial read errors and increased power consumption~\cite{demin2020sneak}. The absence of gating elements and the fully connected structure of the crossbar cause currents to flow preferentially through low-resistance cells~\cite{cassuto2013sneak}, inevitably creating a parallel resistance network that distorts readouts and causes high-resistance state (HRS) to be misinterpreted as low-resistance state (LRS)~\cite{gul2019addressing}. Moreover, the magnitude and distribution of these parasitic currents are highly data-dependent, varying dynamically with the stored information~\cite{tang2018comprehensive} or data patterns. Analyzing sneak path currents during the pre-design and ongoing design phases is crucial, as their data-dependent variability poses significant reliability challenges~\cite{liang2010size}. Without precise estimation of these parasitic currents, it becomes difficult to ensure stable performance in terms of noise margin, functional yield, and maximum achievable array size~\cite{chen2013comprehensive}. 

Some research has focused on mitigating this issue through various strategies. Architectural approaches often incorporate gating elements, such as diodes, transistors, or additional memristors, to suppress unwanted current flow, although at the cost of reduced array density. Alternatively, gateless techniques leverage multi-point read schemes, coding strategies, or logical operations to achieve sneak path free reading~\cite{zidan2013memristor,joshi2021sneak,manem2010design,linn2010complementary,shi2020research,kannan2013sneak}. However, the above studies only limit to the individual aspects such as data patterns, connection configurations, read voltages, or array sizes, rather than capture the combined influence of all key parameters. Consequently, a unified analytical framework that comprehensively models sneak path behavior across multiple interacting factors remains largely unexplored. In addition, recent Computer-Aided Design (CAD) frameworks such as CiMLoop \cite{andrulis2024cimloop}, DNN+NeuroSim \cite{luo2019mlp+}, MNSIM \cite{zhu2023mnsim}, and CrossSim \cite{maram2023deep} have advanced the modeling and co-design of IMC architectures.
For example, CiMLoop enables full-stack exploration from device to architecture; DNN+NeuroSim provides circuit to algorithm benchmarking for SRAM and emerging non-volatile memories; MNSIM  supports hierarchical modeling of digital and analog IMC architectures; and CrossSim simulates ReRAM-based neural training under realistic device nonidealities.
However, these tools neither consider sneak path effects nor employ purely passive memristor arrays; instead, they depend on selector or switching devices to mitigate interference, which substantially limits memory density. This highlights the need for a comprehensive analytical framework dedicated to modeling sneak path phenomena in high-density passive crossbar arrays.

A multi-parameter closed-form analytical model is proposed in this paper to provide a unified framework for evaluating the combined influence of critical design parameters, including device conductance (ON/OFF ratio of memristor), array size, read voltage, interconnect resistance, data patterns, and connection strategies. The proposed approach enables accurate pre-design/real-time estimations of key performance indicators such as functional yield, noise margin, and maximum feasible array size, thereby supporting the development of robust and predictable memristor-based architectures. Compared with conventional SPICE simulations, the proposed closed-form model offers several advantages: (1) Speed and efficiency: it provides immediate analytical results, whereas SPICE requires extensive computation, particularly for large arrays with numerous parasitic components; (2) Scalability: the closed-form model can efficiently handle analyses involving thousands or millions of devices, while SPICE simulations slow down significantly as system size increases; (3) Analytical clarity: unlike SPICE, which produces purely numerical outputs, the closed-form approach reveals explicit relationships among parameters such as resistance, voltage, and current; (4) Design optimization: it allows designers to quickly explore trade-offs in parameters like array size and bias voltage without iterative simulations; and (5) Integration with design tools: the derived equations can be easily incorporated into CAD workflows for rapid performance estimation during schematic, layout or system-level design, whereas SPICE models are less adaptable for high-level design automation. This paper will make the following contributions: 
\begin{itemize}
   \item The dependence of the sneak path current in the memristor crossbar array is analyzed for important circuit parameters. 
   \item A closed-form model is proposed for the accurate and efficient estimation of the sneak path current in memristor arrays, and its accuracy and run-time performance are validated against SPICE simulations.
   \item The sensitivity and noise margins of the sneak path current in the memristor crossbar array are also modeled and deeply analyzed. 
   \item The closed-form model is further compared with the state-of-the-art.
\end{itemize}

This paper is organized into the following sections. In Section~\ref{sec:Methodology}, models and parameters for the memristor, the interconnect network, and closed-form expressions are presented along with the connection mechanism in array configuration. The results and discussions for the sneak path current, noise margin, the effect of interconnect resistance, sensitivity analysis, and model validation are presented in Section~\ref{sec:Results and Discussion}. Performance and property comparison with the state-of-the-art is provided in Section~\ref{sec:Comparison with State-of-The-Art}. Finally, this paper is concluded in Section~\ref{sec:conclusion}.
%\end{enumerate}

%\begin{enumerate}
%    \item Speed and efficiency: it provides immediate analytical results, whereas SPICE requires extensive computation, particularly for large arrays with numerous parasitic components. 
%    \item Scalability: the closed-form model can efficiently handle analyses involving thousands or millions of devices, while SPICE simulations slow down significantly as system size increases. \item Analytical clarity: unlike SPICE, which produces purely numerical outputs, the closed-form approach reveals explicit relationships among parameters such as resistance, voltage, and current. 
%    \item Design optimization: it allows designers to quickly explore trade-offs in parameters like array size and bias voltage without iterative simulations. 
%    \item integration with design tools: the derived equations can be easily incorporated into EDA workflows for rapid performance estimation during schematic, layout, or system-level layout, whereas SPICE models are less adaptable for high-level design automation.
%\end{enumerate}

%Additionally, this study develops closed-form analytical equations to accurately estimate sneak path currents based on these integral parameters. Such an analytical model is crucial for reliable pre-design evaluation of the performance and effectiveness of any memristor-based crossbar array, thereby supporting the development of more robust and predictable memristor-based memory architectures.

\section{Methodology}
\label{sec:Methodology}

\subsection{Design Considerations}
The foundation of this analysis is the nanoscale crossbar memristor model presented in \cite{robinett2007demultiplexers} \textcolor{black}{based on HP Technology}, where the device current is defined in the following equation:
\begin{equation}
I = K \sinh(\alpha V)
\end{equation}
where $\alpha$ indicates the degree of nonlinearity in the device. It indicates the sensitive change of device characteristics due to a unit change in voltage. I and V represent the current and voltage, respectively. The parameter $K$ is analogous to the electrical conductance of a linear resistance.

\begin{figure*}
    \centering
    \includegraphics[width=1\linewidth]
    {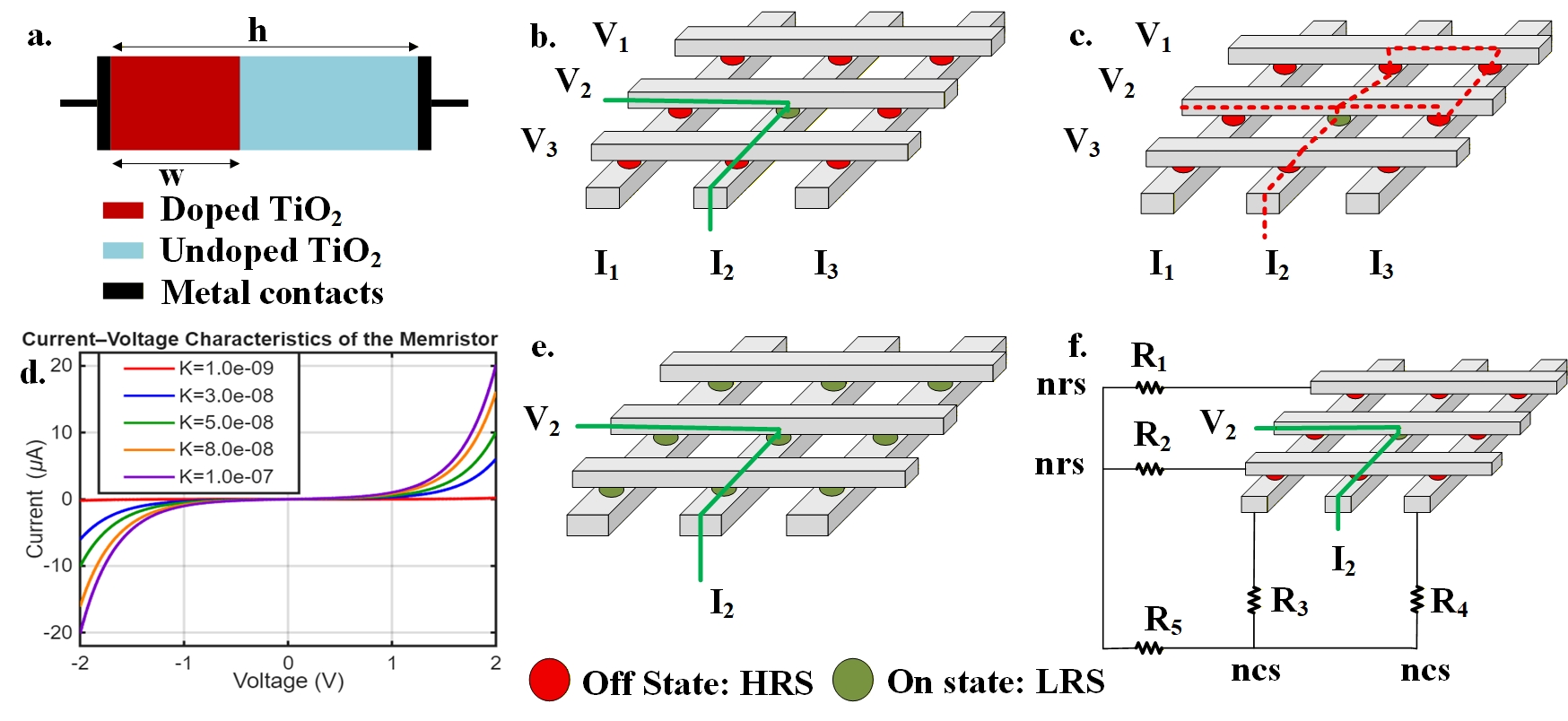}
    \caption{(a) Structure of the linear dopant drift memristor model. h represents the total thickness of the device. w represents the thickness of the doped region; (d) Current versus voltage characteristic for the memristor \cite{robinett2007demultiplexers}. A $3 \times 3$ memristive crossbar, where V\textsubscript{dd} = V\textsubscript{2} is the voltage applied
in row 2, and current sensed from column 2: (b) the crossbar without sneak paths where the desired current path is marked in green line, (c) presence of a sneak path which is marked in red line, (e) FRC connection strategy with `All Ones' (LRS) data pattern, (f) GRC connection strategy with `All Zeros' (HRS) data pattern. Here, in GRC connection strategy, shorted row nodes are indicated as "nrs" and shorted column nodes are indicated as "ncs".}
    \label{fig: Fig1}
\end{figure*}

In this configuration, a TiO\textsubscript{2} layer for a total thickness of the device, \({h} \leq 10\,\text{nm} \), is sandwiched between two platinum electrodes. The oxide layer has two constituent layers: a doped region containing oxygen vacancies and an undoped region composed of pure TiO\textsubscript{2}, as shown in Fig.~\ref{fig: Fig1} (a). This work is conducted using a parameter value of $\alpha$ = 3 \cite{strukov2008missing} and current versus voltage characteristic for the memristor is shown in Fig.~\ref{fig: Fig1} (d) \cite{robinett2007demultiplexers}.

The current method presented in this paper employs the constant K\textsubscript{on} and K\textsubscript{off} as the ON and OFF states, respectively. To comprehensively assess the influence of sneak path currents, two representative data patterns are simulated: the `All Zeros' pattern (best case scenario, minimum sneak path current effect, all the memristors are in the OFF state and the sneak paths are made of OFF resistances in series); and the `All Ones' pattern (worst case scenario, maximum sneak path current effect). Grounded unselected rows and columns has been explored as a method to mitigate sneak path currents. Grounded rows and columns of unselected cells in an array provides alternating paths for leakage currents to flow instead of the sensing circuit. However, part of the sneak path current still flows through the sense circuit. Four terminal biasing configurations are analyzed: Floating Rows and Columns (FRC), Grounded Rows with Floating Columns (GRFC), Floating Rows with Grounded Columns (FRGC), and Grounded Rows and Columns (GRC), to examine the impact of different connection strategies on the overall array behavior. The grounded rows and columns are implemented using \( 0.01\,\Omega \) resistors (near zero), with the corresponding nodes labeled as node row shorted and node column shorted, respectively.

% \begin{figure*}
  % \centering
    %\includegraphics[width=\columnwidth]{Figures/Fig2.jpg}
   %  \subfloat{\includegraphics[width=1\linewidth]{Figures/Fig10.jpg}} \\
    % \subfloat{\includegraphics[height= 2in,width=0.50\linewidth]{Figures/Fig11.jpg}}
    % \subfloat{\includegraphics[width=0.40\linewidth]{Figures/Fig12.jpg}}
   % \caption{(a) Metal layers M1-M12 are used for interconnect within the cell, (b) The sketch of Cu for the interconnect material, (c) Top view of the unit-cell memristor crossbar array. }
  %  \label{fig: Fig2}
%\end{figure*}

\begin{figure*}
   \centering
    \includegraphics[height=1.8in, width=1.9\columnwidth]
    {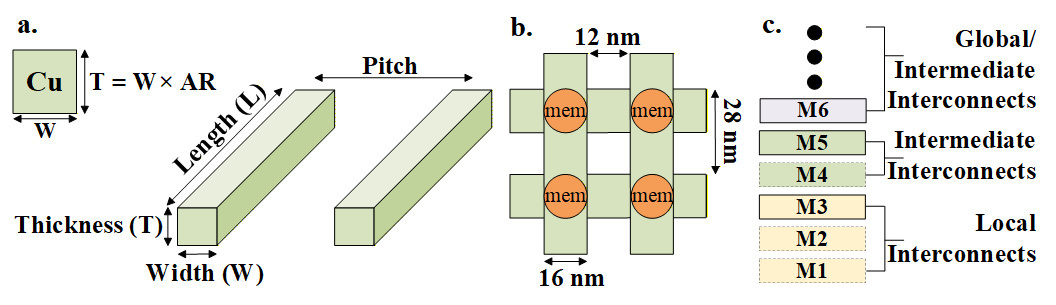}
    \caption{(a) Cu interconnect geometry, with barrier thickness taken into account; (b) Top view of the unit-cell memristor crossbar array for M5 with 1.4nm technology; (c) Metal layers are used for interconnects within cells.}
    \label{fig: Fig2}
\end{figure*}

 Fig.~\ref{fig: Fig1} presents an abstract depiction of the sneak path phenomenon in the crossbar array, along with schematic representations to illustrate the different combinations of data patterns and connection strategies. The device at the center of the array is the ``target device," and the devices sharing a line with the target are called ``half-selected devices". To read the ``target device", the selected row is pulled up to V\textsubscript{dd} and the selected column is grounded through a resistor. Sneak path occurs when current flows from V\textsubscript{dd} to ground through unselected devices. Therefore, it affects the accuracy of the read operation by degrading the voltage and current delivered to the device.
The crossbar interconnect network is modeled at the unit-cell level to accurately capture parasitic effects, incorporating line resistance, stray capacitance to ground, and fringing capacitance between adjacent lines. In this work, the interconnect design dimensions follow the 1.4nm technology node, offering concrete guidance for both academic researchers and industry engineers \cite{pei2025interconnect,pei2024ultra,liu2023cfet,pei2023technology,liu2024future,liu2023cfetted}. Generally, three interconnect layers M3, M5, and M6 with different geometries are chosen to represent interconnects at local, intermediate, and global levels, respectively, as illustrated in Fig.~\ref{fig: Fig2}. M3 can be used for local interconnects. M5 is for the intermediate-level interconnects. M6 and above are for intermediate- and global-level interconnects. Here, the interconnect geometry is characterized by the width (W), thickness (T), and pitch (P), as illustrated in Fig.~\ref{fig: Fig2} (a). The aspect ratio (AR) is defined as the ratio of T to W, while P denotes the center-to-center spacing between adjacent interconnects. The unit-cell memristor array, annotated with the corresponding values of W and P, is depicted in Fig.~\ref{fig: Fig2} (b). Fig.~\ref{fig: Fig2} (c) presents the utility of metal layers for different interconnects. The selected interconnect metal layers provide different parasitic resistance R\textsubscript{line} and capacitance C\textsubscript{line}. The corresponding values of interconnect parameters are listed in Table~\ref{table:imec_a14}.

\begin{table}[!ht]
    \centering
    \caption{Interconnect Geometry, Resistance, and Capacitance}
    \label{table:imec_a14}
    \begin{adjustbox}{width=\columnwidth}								   
    \begin{tabular}{|c|c|c|c|c|c|}
    \hline
        Metal & W(nm) & P(nm) & T(nm) & R\textsubscript{line}($\Omega$) & C\textsubscript{line}(fF) \\ \hline
        M3 & 16 & 28 & 49 & 3.122 & 3.871e-3 \\ \hline
        M5 & 16 & 28 & 28 & 5.869 & 2.871e-3 \\ \hline
        M6 & 40 & 80 & 80 & 0.7396 & 1.020e-2 \\ \hline
    \end{tabular}
\end{adjustbox}   
\end{table}

\begin{table}[!ht]
\centering
\caption{Boundary Conditions for Closed-form Models}
\label{table:closed-form}
\begin{adjustbox}{width=\columnwidth}				  
\begin{tabular}{|c|c|c|c|c|c|c|c|c|}
\hline
        Para. & Boundary & Para. & Boundary & Para. & Boundary  \\ \hline
        Size& $4$×$4$:$64$×$64$ & $K_{on}$ & 1e-9:1e-7 & $V_{dd}$ & 1V:3V  \\ \hline
\end{tabular}
\end{adjustbox}
\end{table}

%\begin{figure*}
   %  \centering
   %  \subfloat[]{\includegraphics[width=0.80\linewidth]{Figures/Fig10.jpg}} \\
    % \subfloat[]{\includegraphics[width=0.40\linewidth]{Figures/Fig11.jpg}}
    % \subfloat[]{\includegraphics[width=0.40\linewidth]{Figures/Fig12.jpg}}
     
   %\caption{(a) Metal layers M1-M12 are used for interconnect within the cell, (b) The sketch of Cu for the interconnect material, (c) Top view of the unit-cell memristor crossbar array. }
  %  \label{fig: Fig2}
% \end{figure*}

%\textcolor{red}{1. What is W, P, H, Rw, and Cw? 2. Where indicating Fig. 2? 3. add descriptions foir Fig.2 b and c. 4. Please make the good positions for Figs. 1 and 2}

\begin{figure*}
     \centering
  {\includegraphics[width=1\linewidth]{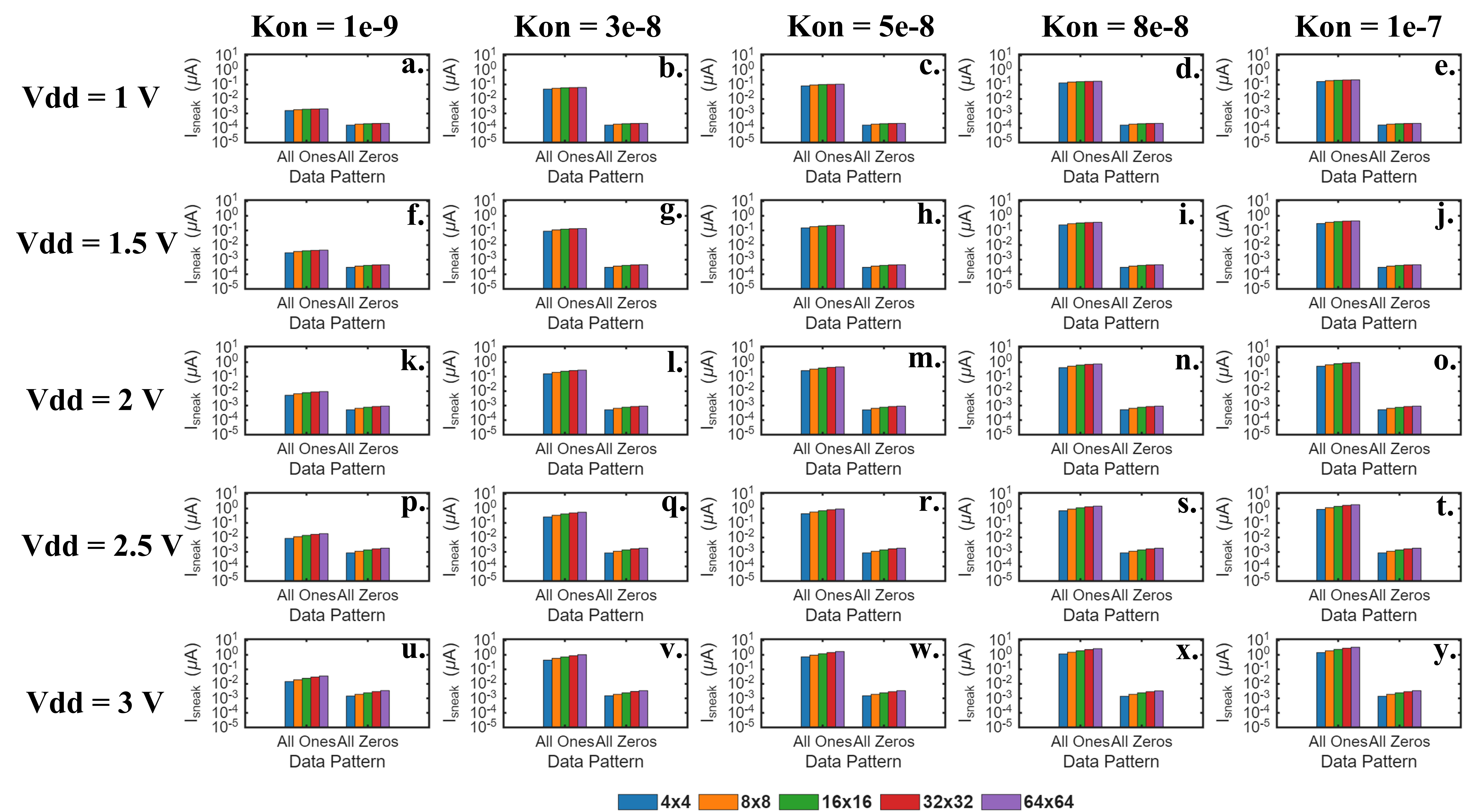}}
     \caption{ Simulation results for sneak path current in terms of array sizes, data patterns (`All Zeros' and `All Ones'), FRC connection strategy, and line resistance M3 (R\textsubscript{line} 
     = \( 3.122\,\Omega \)) where the read voltages are incremented along the rows and ON/OFF ratio is incremented along the columns of the subplots.
     %(a) K\textsubscript{on}= \(10^{-7} \),  V\textsubscript{dd} = 3 V; (b)  K\textsubscript{on}= \(10^{-7} \),  V\textsubscript{dd} = 1 V; (c) K\textsubscript{on}= \(10^{-9} \) , V\textsubscript{dd} = 3 V;  (d) K\textsubscript{on}= \(10^{-9} \), V\textsubscript{dd} = 1 V.}
    The rise in sneak path current with respect to the increment of the ON/OFF ratio along the columns as K\textsubscript{on} = [1e-9, 3e-8, 5e-8, 8e-8, 1e-7] for a specific read voltage: (a\textasciitilde e) V\textsubscript{dd} = 1.0 V; (f\textasciitilde j) V\textsubscript{dd} = 1.5 V; (k\textasciitilde o) V\textsubscript{dd} = 2.0 V; (p\textasciitilde t) V\textsubscript{dd} = 2.5 V; (u\textasciitilde y) V\textsubscript{dd} = 3 V. Noticeably, for `All Ones' pattern the sneak path current increases rapidly for the increments of read voltage, ON/OFF ratio, array size compared to the minimal increment of All Zeros data pattern.}

     \label{fig:Fig3}
\end{figure*}

\begin{table*}[!ht]
\centering
\caption{Coefficients of Closed-form Expression for M3}
\label{tab:results1_1_M3}
\begin{tabular}{|c|c|c|c|c|c|c|c|c|}
\hline
\textbf{Coeff.}  & \textbf{All 1s FRC} & \textbf{All 1s GRFC} & \textbf{All 1s FRGC} & \textbf{All 1s GRC} & \textbf{All 0s FRC} & \textbf{All 0s GRFC} & \textbf{All 0s FRGC} & \textbf{All 0s GRC} \\
\hline
\textbf{$C_{1}$}  & -2.765766e-04 & -5.271610e-04 & -3.118053e-05 & -3.422062e-05 & -2.768951e-04 & -5.173225e-04 & 6.728880e-08 & 7.209101e-08 \\
\hline
\textbf{$C_{2}$}  & -3.552098e-05 & -7.224109e-04 & -9.791516e-04 & -9.441148e-04 & -2.337750e-05 & -7.911723e-05 & -1.496197e-04 & -1.496397e-04 \\
\hline
\textbf{$C_{3}$}  & 4.599539e-03 & 2.568702e-03 & -3.942867e-03 & -3.851644e-03 & 4.602914e-03 & 4.479660e-03 & -8.406540e-04 & -8.401264e-04 \\
\hline
\textbf{$C_{4}$}  & 1.722779e-02 & 3.588961e-02 & -1.070919e-02 & -9.972648e-03 & 1.747226e-02 & 4.457329e-02 & -1.435455e-03 & -1.437187e-03 \\
\hline
\textbf{$C_{5}$}  & -4.296973e-04 & -9.912720e-03 & -3.415353e-02 & -3.305335e-02 & -1.110902e-03 & -1.103802e-03 & -2.031960e-03 & -2.030510e-03 \\
\hline
\textbf{$C_{6}$}  & -1.275372e-03 & -2.229373e-02 & -1.746779e-01 & -1.730254e-01 & -1.192496e-03 & -3.191236e-03 & -6.087109e-03 & -6.086547e-03 \\
\hline
\textbf{$C_{7}$}  & 9.867175e-01 & 6.749072e-01 & -2.319980e-02 & 1.489677e-02 & -3.901237e-02 & -3.450177e-02 & -6.308356e-02 & -6.303070e-02 \\
\hline
\textbf{$C_{8}$}  & -1.056307e-01 & -1.838965e-01 & -3.936425e-01 & -3.919864e-01 & -1.029701e-01 & -1.556039e-01 & -2.579691e-02 & -2.578021e-02 \\
\hline
\textbf{$C_{9}$}  & 1.529703e+00 & 1.871899e+00 & 1.029943e+00 & 1.054458e+00 & 1.521661e+00 & 2.102373e+00 & 2.995661e+00 & 2.995596e+00 \\
\hline
\textbf{$C_{10}$}  & -1.154712e+00 & -3.592643e+00 & -8.795638e+00 & -8.468067e+00 & -2.441334e+01 & -2.416504e+01 & -2.428478e+01 & -2.428423e+01 \\
\hline
\end{tabular}
\end{table*}

\begin{table*}[!ht]
\centering
\caption{Coefficients of Closed-form Expression for M5}
\label{tab:results1_2_M5}
\begin{tabular}{|c|c|c|c|c|c|c|c|c|}
\hline
\textbf{Coeff.}  & \textbf{All 1s FRC} & \textbf{All 1s GRFC} & \textbf{All 1s FRGC} & \textbf{All 1s GRC} & \textbf{All 0s FRC} & \textbf{All 0s GRFC} & \textbf{All 0s FRGC} & \textbf{All 0s GRC} \\
\hline
\textbf{$C_{1}$}  & -2.764303e-04 & -5.369728e-04 & -3.317372e-05 & -3.607482e-05 & -2.795334e-04 & -5.162575e-04 & 1.109504e-06 & 1.100590e-06 \\
\hline
\textbf{$C_{2}$}  & -5.936502e-05 & -1.011910e-03 & -1.426496e-03 & -1.405009e-03 & -4.490970e-05 & -1.323556e-04 & -2.524449e-04 & -2.524109e-04 \\
\hline
\textbf{$C_{3}$}  & 4.487461e-03 & 1.771752e-03 & -5.595220e-03 & -5.549250e-03 & 4.561308e-03 & 4.208091e-03 & -1.425890e-03 & -1.425737e-03 \\
\hline
\textbf{$C_{4}$}  & 1.695349e-02 & 3.218355e-02 & -1.690781e-02 & -1.636382e-02 & 1.732497e-02 & 4.394577e-02 & -2.493809e-03 & -2.492790e-03 \\
\hline
\textbf{$C_{5}$}  & -7.048080e-04 & -1.162975e-02 & -3.488875e-02 & -3.399748e-02 & -2.417057e-04 & -1.852363e-03 & -3.402301e-03 & -3.402806e-03 \\
\hline
\textbf{$C_{6}$}  & -2.179162e-03 & -2.706075e-02 & -1.823875e-01 & -1.810900e-01 & -1.532339e-03 & -5.523961e-03 & -1.053107e-02 & -1.053004e-02 \\
\hline
\textbf{$C_{7}$}  & 9.783429e-01 & 6.228614e-01 & -3.333252e-02 & -2.208753e-03 & -5.741777e-03 & -5.769302e-02 & -1.051583e-01 & -1.051793e-01 \\
\hline
\textbf{$C_{8}$}  & -1.081555e-01 & -1.901170e-01 & -3.993790e-01 & -3.982103e-01 & -1.062961e-01 & -1.624248e-01 & -4.035391e-02 & -4.034864e-02 \\
\hline
\textbf{$C_{9}$}  & 1.524105e+00 & 1.817728e+00 & 9.227569e-01 & 9.431813e-01 & 1.527636e+00 & 2.088758e+00 & 2.976207e+00 & 2.976206e+00 \\
\hline
\textbf{$C_{10}$}  & -1.225103e+00 & -4.002326e+00 & -8.767657e+00 & -8.499305e+00 & -2.411070e+01 & -2.436257e+01 & -2.464774e+01 & -2.464794e+01 \\
\hline
\end{tabular}
\end{table*}

\begin{table*}[!ht]
\centering
\caption{Coefficients of Closed-form Expression for M6}
\label{tab:results1_3_M6}
\begin{tabular}{|c|c|c|c|c|c|c|c|c|}
\hline
\textbf{Coeff.}  & \textbf{All 1s FRC} & \textbf{All 1s GRFC} & \textbf{All 1s FRGC} & \textbf{All 1s GRC} & \textbf{All 0s FRC} & \textbf{All 0s GRFC} & \textbf{All 0s FRGC} & \textbf{All 0s GRC} \\
\hline
\textbf{$C_{1}$}  & -2.765330e-04 & -5.136305e-04 & -1.469021e-05 & -1.697019e-05 & -2.744689e-04 & -5.168549e-04 & -1.772653e-07 & -1.834111e-07 \\
\hline
\textbf{$C_{2}$}  & -9.275605e-06 & -3.903261e-04 & -3.335456e-04 & -2.664594e-04 & -1.905130e-05 & -2.198834e-05 & -3.958725e-05 & -3.958864e-05 \\
\hline
\textbf{$C_{3}$}  & 4.723620e-03 & 3.578035e-03 & -1.398252e-03 & -1.188272e-03 & 4.644600e-03 & 4.760412e-03 & -2.236606e-04 & -2.236878e-04 \\
\hline
\textbf{$C_{4}$}  & 1.751128e-02 & 3.976450e-02 & -3.205943e-03 & -2.113479e-03 & 1.729086e-02 & 4.514727e-02 & -3.611497e-04 & -3.607864e-04 \\
\hline
\textbf{$C_{5}$}  & -1.186960e-04 & -7.747277e-03 & -3.295507e-02 & -3.143524e-02 & -6.912710e-04 & -2.288891e-04 & -5.495305e-04 & -5.487445e-04 \\
\hline
\textbf{$C_{6}$}  & -3.378056e-04 & -1.659931e-02 & -1.633178e-01 & -1.609204e-01 & -5.466582e-05 & -7.376962e-04 & -1.586451e-03 & -1.588074e-03 \\
\hline
\textbf{$C_{7}$}  & 9.963053e-01 & 7.411125e-01 & -5.549090e-03 & 4.635883e-02 & -2.498576e-02 & -6.894301e-03 & -1.715344e-02 & -1.712207e-02 \\
\hline
\textbf{$C_{8}$}  & -1.029360e-01 & -1.751892e-01 & -3.814526e-01 & -3.785520e-01 & -1.014793e-01 & -1.482829e-01 & -1.070043e-02 & -1.069628e-02 \\
\hline
\textbf{$C_{9}$}  & 1.534969e+00 & 1.929448e+00 & 1.164724e+00 & 1.196345e+00 & 1.538061e+00 & 2.116129e+00 & 3.012977e+00 & 3.012930e+00 \\
\hline
\textbf{$C_{10}$}  & -1.073100e+00 & -3.058583e+00 & -8.790361e+00 & -8.344802e+00 & -2.429351e+01 & -2.392658e+01 & -2.388512e+01 & -2.388481e+01 \\
\hline
\end{tabular}
\end{table*}

\subsection{Closed-form Model}
Closed-form expressions for the sneak current are presented corresponding to different input variables, e.g., Size (Array Size), K\textsubscript{on} (ON/OFF ratio), and V\textsubscript{dd} (Read Voltage), as Equation~\eqref{eq:closedformexpression}. The boundaries for closed-form equation fitting are selected to encompass the full range of practical circuit biasing conditions and fabrication considerations. The corresponding ranges are listed in Table~\ref{table:closed-form}. As shown in Equation~\eqref{eq:closedformexpression}, the corresponding coefficients are summarized in Tables~\ref{tab:results1_1_M3},~\ref{tab:results1_2_M5},~\ref{tab:results1_3_M6}. 
%Three metal layers M3, M5, and M6 of 1.4nm with different geometries are chosen to represent interconnects at local, intermediate, and global levels, respectively, with M1 and M2 primarily employed for internal interconnect with cell. Metal layer M3 can be used for local interconnect within subarrays and M4 and M5 are for the intermediate-level interconnect. Metal layers M6 and above are for intermediate- and global-level interconnects Table~\ref{table:imec_a14} \cite{pei2025interconnect,pei2024ultra,liu2023cfet,pei2023technology,liu2024future,liu2023cfetted}.

%%\begin{multline}
\begin{align}
\label{eq:closedformexpression}
%&I_{sneak} = exp(C_{1}\cdot Size^{2}+C_{2} \cdot Size\cdot ln(K_{on}) \notag \\& +C_{3}\cdot Size\cdot R_{cb} +C_{4} \cdot Size\cdot V_{dd}+C_{5}\cdot Size \notag \\&+C_{6}\cdot ln(K_{on})^{2} +C_{7} \cdot ln(K_{on}) \cdot R_{cb} +C_{8}\cdot ln(K_{on})\cdot V_{dd} \notag \\& +C_{9}\cdot ln(K_{on}) +C_{10}\cdot R_{cb}^{2} +C_{11}\cdot R_{cb}\cdot V_{dd} \notag \\&+C_{12}\cdot R_{cb} +C_{13}\cdot V_{dd}^{2}+C_{14}\cdot V_{dd}+C_{15})
&I_{sneak} = exp(C_{1}\cdot Size^{2}+C_{2}\cdot Size\cdot ln(K_{on})\notag \\&+C_{3}\cdot Size\cdot V_{dd}+C_{4}\cdot Size+C_{5}\cdot ln(K_{on})^{2}\notag \\&+C_{6}\cdot ln(K_{on})\cdot V_{dd}+C_{7}\cdot ln(K_{on})+C_{8}\cdot V_{dd}^{2} \textbf{}\notag \\&+C_{9}\cdot V_{dd}+C_{10})
\end{align}

\begin{table*}[!ht]
    \centering

    \caption{Sneak Path Current Model Validation: Simulation Results, Errors, and Normalized Run-time Improvement}
    \label{tab:results2}
    \begin{tabular}{|c|c|c|c|c|c|c|c|c|c|c|c|}

    \hline

\multirow{2}{*}{Pattern} &
\multirow{2}{*}{Strategy} &
\multirow{2}{*}{Metal} &
\multicolumn{3}{c|}{Size:8×8, Kon:3e-8, Vdd:1.5V} &
\multicolumn{3}{c|}{Size:16×16, Kon:5e-8, Vdd:2V} &
\multicolumn{3}{c|}{Size:32×32, Kon:8e-8, Vdd:2.5V} \\
\cline{4-12}
~ & ~ & ~ & Simulation(A) & Error & Run-time & Simulation(A) & Error & Run-time & Simulation(A) & Error & Run-time\\ \hline
        All 1s & FRC & M3 & 1.089e-07 & -7.017\% & 1916× & 3.896e-07 & -6.257\% & 835× & 1.312e-06 & 2.163\% & 2030×  \\ \hline
        All 1s & FRC & M5 & 1.089e-07 & -6.895\% & 768× & 3.895e-07 & -6.213\% & 159× & 1.307e-06 & 2.052\% & 4784×  \\ \hline
        All 1s & FRC & M6 & 1.089e-07 & -7.151\% & 1132× & 3.897e-07 & -6.319\% & 1466× & 1.316e-06 & 2.325\% & 1821×  \\ \hline
        All 1s & GRFC & M3 & 3.608e-07 & -6.321\% & 1198× & 1.866e-06 & -7.207\% & 62× & 8.210e-06 & 7.760\% & 1777×  \\ \hline
        All 1s & GRFC & M5 & 3.608e-07 & -5.789\% & 1103× & 1.861e-06 & -7.205\% & 2048× & 7.935e-06 & 7.308\% & 1464×  \\ \hline
        All 1s & GRFC & M6 & 3.609e-07 & -6.993\% & 690× & 1.870e-06 & -7.405\% & 1355× & 8.479e-06 & 8.717\% & 1023×  \\ \hline
        All 1s & FRGC & M3 & 1.312e-06 & 10.573\% & 730× & 8.314e-06 & -4.316\% & 542× & 2.969e-05 & -3.853\% & 1618×  \\ \hline
        All 1s & FRGC & M5 & 1.312e-06 & 10.857\% & 1378× & 8.202e-06 & -4.910\% & 137× & 2.669e-05 & -2.296\% & 1561×  \\ \hline
        All 1s & FRGC & M6 & 1.313e-06 & 10.043\% & 43× & 8.415e-06 & -3.978\% & 888× & 3.336e-05 & -4.605\% & 1876×  \\ \hline
        All 1s & GRC & M3 & 1.313e-06 & 10.260\% & 10× & 8.331e-06 & -4.263\% & 1024× & 3.014e-05 & -4.124\% & 1230×  \\ \hline
        All 1s & GRC & M5 & 1.312e-06 & 10.618\% & 172× & 8.218e-06 & -4.857\% & 779× & 2.703e-05 & -2.529\% & 1694×  \\ \hline
        All 1s & GRC & M6 & 1.313e-06 & 9.584\% & 1091× & 8.433e-06 & -3.967\% & 2426× & 3.401e-05 & -4.837\% & 1698×  \\ \hline
        All 0s & FRC & M3 & 3.630e-10 & -6.942\% & 826× & 7.800e-10 & -6.381\% & 305× & 1.640e-09 & 2.166\% & 1539×  \\ \hline
        All 0s & FRC & M5 & 3.630e-10 & -7.075\% & 746× & 7.800e-10 & -6.337\% & 1392× & 1.630e-09 & 2.753\% & 1646×  \\ \hline
        All 0s & FRC & M6 & 3.630e-10 & -6.964\% & 214× & 7.800e-10 & -6.286\% & 1382× & 1.650e-09 & 1.980\% & 1229×  \\ \hline
        All 0s & GRFC & M3 & 1.208e-09 & -8.582\% & 123× & 3.810e-09 & -7.350\% & 1269× & 1.162e-08 & 8.837\% & 1385×  \\ \hline
        All 0s & GRFC & M5 & 1.208e-09 & -8.254\% & 977× & 3.810e-09 & -7.298\% & 1489× & 1.151e-08 & 8.491\% & 1672×  \\ \hline
        All 0s & GRFC & M6 & 1.208e-09 & -8.930\% & 684× & 3.820e-09 & -7.701\% & 319× & 1.171e-08 & 9.336\% & 1873×  \\ \hline
        All 0s & FRGC & M3 & 4.499e-09 & 1.050\% & 965× & 2.013e-08 & 0.573\% & 600× & 8.839e-08 & -1.321\% & 1827×  \\ \hline
        All 0s & FRGC & M5 & 4.499e-09 & 1.765\% & 1263× & 2.011e-08 & 0.837\% & 1352× & 8.686e-08 & -2.087\% & 1814×  \\ \hline
        All 0s & FRGC & M6 & 4.500e-09 & 0.283\% & 774× & 2.016e-08 & 0.167\% & 1615× & 8.978e-08 & -0.420\% & 1167×  \\ \hline
        All 0s & GRC & M3 & 4.499e-09 & 1.049\% & 806× & 2.013e-08 & 0.571\% & 1176× & 8.839e-08 & -1.321\% & 1748×  \\ \hline
        All 0s & GRC & M5 & 4.499e-09 & 1.765\% & 954× & 2.011e-08 & 0.837\% & 930× & 8.686e-08 & -2.085\% & 945×  \\ \hline
        All 0s & GRC & M6 & 4.500e-09 & 0.281\% & 1078× & 2.016e-08 & 0.166\% & 1455× & 8.978e-08 & -0.421\% & 800×  \\ \hline
    \end{tabular}
\end{table*}

\section{Results and Discussion}
\label{sec:Results and Discussion}
\subsection{ Sneak Path Current Analysis}

The sneak path effect is examined for different memristor crossbar array sizes; the analysis also includes a range of device parameters and data patterns. The magnitude of the sneak path current and the corresponding array size have a strong correlation. As the size of the array increases, the number of potential parallel sneak paths increases, leading to an increase in the current value of the sneak path for a given memristor resistance at a given input voltage value. Fig.~\ref{fig:Fig3} demonstrates the scalability challenge, where dense arrays face exponentially worse sneak path issues. The effect of the data pattern in the array is also shown in Fig.~\ref{fig:Fig3}, where all memristors are in a LRS, creating the maximum number of low-resistance parallel paths. Consequently, the ``All Ones" pattern consistently results in the highest sneak path current across all array sizes and parameter sets. Similarly, all memristors with a HRS present the highest overall resistance to current flow. As a result, the ``All Zeros" pattern consistently yields the lowest sneak path currents in all tested conditions.
Additionally, Fig.~\ref{fig:Fig3} illustrates how different device parameters can either mitigate or aggravate the sneak path problem. Considering the FRC connection strategy with M3 as the interconnect metal, the dependence of K\textsubscript{on} value and V\textsubscript{dd} value highly influences the sneak path current as shown in Fig.~\ref{fig:Fig3} (a-y). As the read voltage (V\textsubscript{dd}) increases from 1 V to 3 V with a step size of 0.5 V, a significant increase in sneak path current can be observed in Fig.~\ref{fig:Fig3} (a, f, k, p, u). The K\textsubscript{on} values are swept while keeping the V\textsubscript{dd} fixed, and the trend of sneak path current is observed across five rows in Fig.~\ref{fig:Fig3}. The first row indicates the sweeping of K\textsubscript{on} with a linear increment for a fixed V\textsubscript{dd}, while the next four rows have similar K\textsubscript{on} sweeping but have different V\textsubscript{dd} values across consecutive rows. As the ON/OFF ratio of memristor gets increased by having K\textsubscript{off} set at 1e-10, and changing K\textsubscript{on} to 1e-9, 3e-8, 5e-8, 8e-8, and 1e-7, the sneak path current also increases, which is observed in Fig.~\ref{fig:Fig3}. Noticeably, the sneak path current is negligible for small array sizes, but increases significantly for large array sizes for both `All Ones' and `All Zeros' data patterns. However, for the `All Zeros' pattern, the increase of sneak path current is significantly lower at a higher ON/OFF ratio as the resistances tend to become ineffective at a higher ON/OFF ratio.
%\textcolor{red}{When the ON-state conductance is high (K\textsubscript{on}= \(10^{-7} \), K\textsubscript{off}= \(10^{-10} \))  and the read voltage is large (V\textsubscript{dd} = 3 V), significant sneak path currents are observed even in smaller arrays [Fig.~\ref{fig:Fig3} (a)]. In contrast, a lower (K\textsubscript{on}= \(10^{-9} \), K\textsubscript{off}= \(10^{-10} \)), representing a less conductive ON-state, results in considerably smaller sneak path currents [Fig.~\ref{fig:Fig3} (c)]. Finally, lowering the read voltage from 3 V to 1 V [Fig.~\ref{fig:Fig3} (b, d)] further reduces the sneak path effect, particularly for the all-zero data pattern, while the sneak path current for the all-one pattern shows a modest decrease compared to Fig.~\ref{fig:Fig3} (a, c).}

\subsection{Noise Margin Analysis}
The allowable voltage levels in a CMOS circuit for logic ZERO should be ideally V\textsubscript{OL}= GND, and for logic ONE, it should be V\textsubscript{OH} = V\textsubscript{dd}. However, some variations in these values are acceptable under specific bounds, which are referred to as `Noise Margin'. The noise margin of a memristor cell is defined as the following equation:
\begin{equation}
Margin_{noise} = V_{one} - V_{zero}
\end{equation}
where $V_{one}$ and $V_{zero}$ denote the voltages of the target cell in the low-resistance and high-resistance states, respectively. For a memristor crossbar array, the voltage values representing logic ONE and ZERO are highly affected due to the sneak path current. The noise margin of a target memristor cell is a strong indicator of whether the cell is operating effectively in an array. A large noise margin means the sensing circuit can easily identify the stored bits despite the sneak path effect and other disturbances. In contrast, a small noise margin usually means that the sensed voltages are too close to distinguish, which makes the cell unreliable and effectively nonfunctional in the array. The sneak path effect on noise margin is indicated by the difference between the sensed voltages corresponding to stored ZEROs and ONEs at the target cell. %The Noise Margin of a memristor cell is defined as the following equation:
%\begin{equation}
%Noise  Margin = V\textsubscript{one} - V\textsubscript{zero}
%\end{equation}

For the analysis of the sneak path effect on the noise margin of the memristor array, a normalized parameter is defined 
\cite{zidan2013memristor}:
\begin{equation}
Margin_{noise} = \frac{Margin_{noise\_array}}{Margin_{noise\_device}}
\end{equation}

Here, Margin\textsubscript{noise\_device} represents the noise margin of a single device without any interference of the sneak path, as the single device does not counter the sneak path issue, and Margin\textsubscript{noise\_array} represents the voltage difference of the ON and OFF states of a target cell while being subjected to operation in a memristor array. The noise margin of a target cell in an array highly depends on the data pattern stored within memory. For our analysis, only the best case (All Zeros) and the worst case (All Ones) data patterns are introduced. 

\begin{figure*}
     \centering
  {\includegraphics[width=1\linewidth]{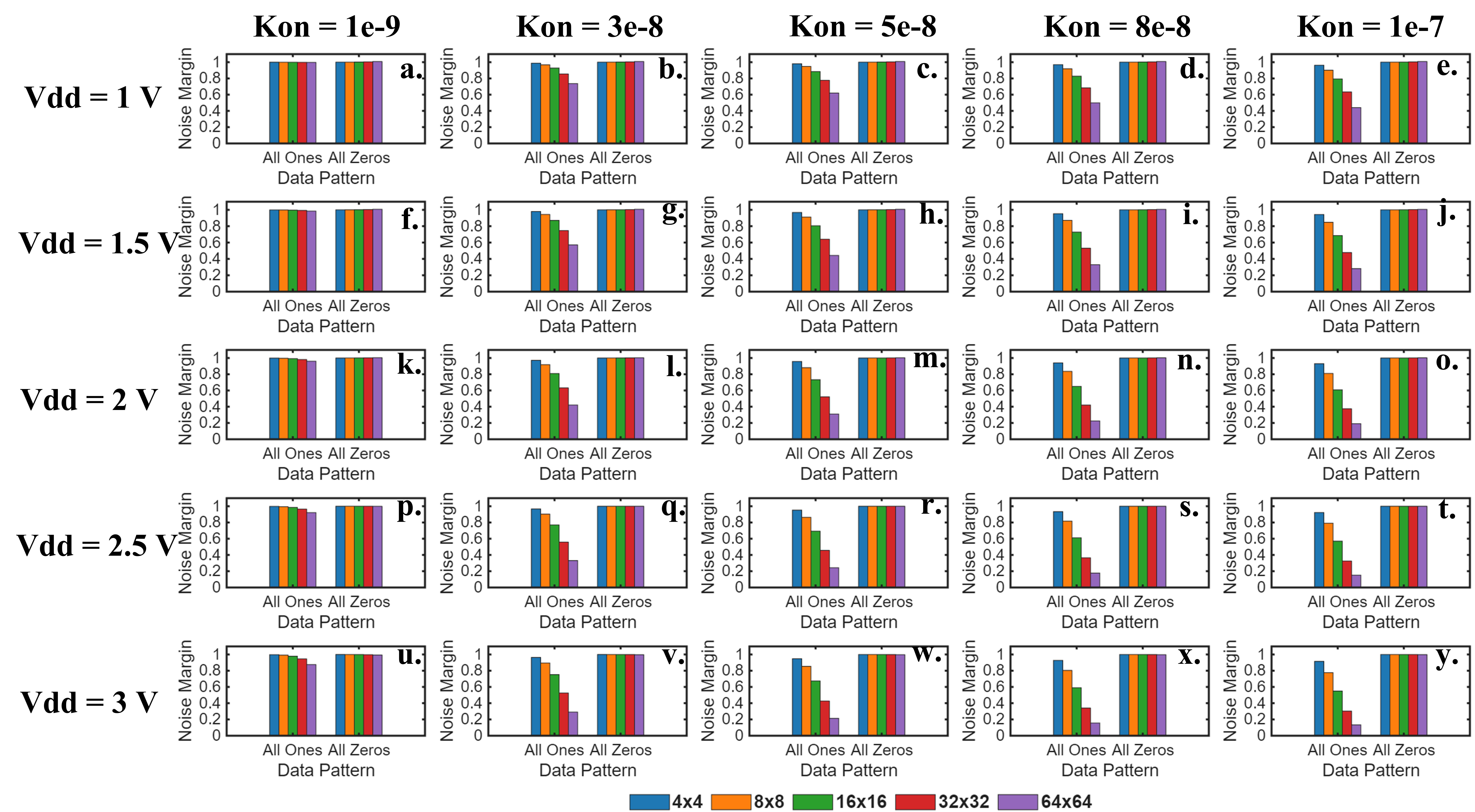}}
     \caption{Simulation results for noise margin versus array sizes, data patterns (`All Zeros' and `All Ones'), FRC connection strategy, and line resistance M3 (R\textsubscript{line} = \( 3.122\,\Omega \)) where the read voltages are incremented along the rows and ON/OFF ratio is incremented along the columns of the subplots.
     %(a) K\textsubscript{on}= \(10^{-7} \),  V\textsubscript{dd} = 3 V; (b)  K\textsubscript{on}= \(10^{-7} \),  V\textsubscript{dd} = 1 V; (c) K\textsubscript{on}= \(10^{-9} \) , V\textsubscript{dd} = 3 V;  (d) K\textsubscript{on}= \(10^{-9} \), V\textsubscript{dd} = 1 V.}
    The fall of noise margin for `All Ones' data pattern with respect to the increment of the ON/OFF ratio along the columns as K\textsubscript{on} = [1e-9, 3e-8, 5e-8, 8e-8, 1e-7] for a specific read voltage: (a\textasciitilde e) V\textsubscript{dd} = 1.0 V; (f\textasciitilde j) V\textsubscript{dd} = 1.5 V; (k\textasciitilde o) V\textsubscript{dd} = 2.0 V; (p\textasciitilde t) V\textsubscript{dd} = 2.5 V; (u\textasciitilde y) V\textsubscript{dd} = 3 V. Noticeably, for `All Ones' pattern, noise margin rapidly decreases with respect to the increments of read voltage, ON/OFF ratio, array size compared to approximately zero decrease for `All Zeros' data pattern.}
     \label{fig:Fig7}
\end{figure*}

Taking M3 as an example, the simulation results in Fig.~\ref{fig:Fig7} demonstrate that at small array sizes the noise margins of both `All Ones' and `All Zeros' are almost similar at lower read voltage, e.g., V\textsubscript{dd}= 1 V and smaller ON/OFF ratio, e.g., K\textsubscript{on}= 1e-9. However, as the read voltage and K\textsubscript{on} increase, the noise margin of larger arrays tends to decrease significantly. This is due to the increase of the sneak path current in larger arrays for high read voltage and high ON/OFF ratio. Finally, from the results, the added resistance of the sneak paths significantly narrows the noise margin and reduces the maximum possible size of a memristor array. It is the limiting factor for array scalability.
 
\subsection{ Interconnect Resistance Effect on Sneak Path Current}

\begin{figure*}
    \centering
    \includegraphics[width=1.5\columnwidth]{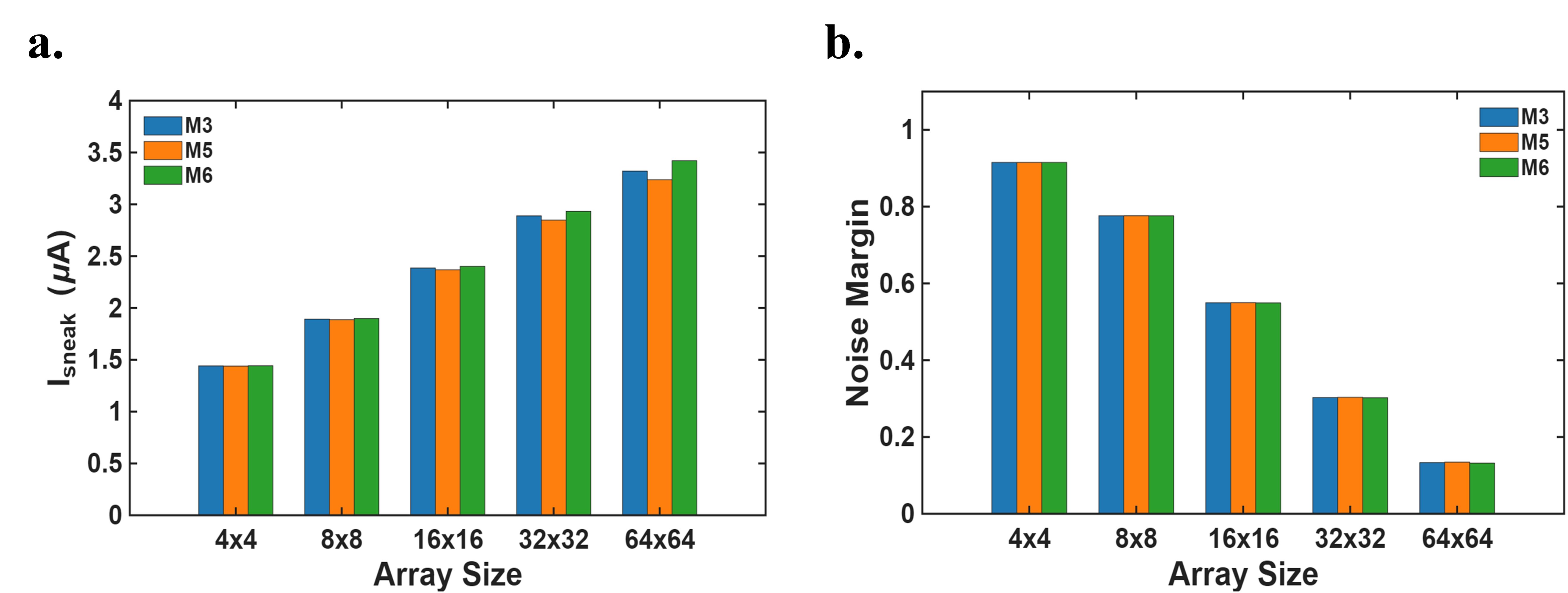}
    \caption{(a) Sneak path current and (b) noise margin of the target cell versus array size for the interconnect layers using the 1.4nm technology. The array adopts ‘All Ones’ data pattern with FRC connection strategy. K\textsubscript{on} is 1e-7. V\textsubscript{dd} is 3 V.  }
    \label{fig: Fig8}
\end{figure*}

The analysis demonstrates that the influence of three line resistances (M3, M5, M6), R\textsubscript{line}, on sneak path currents becomes more pronounced as the array size scales, though its relative impact remains secondary to the inherent sneak path phenomenon. In small arrays, the line resistance is negligible compared to the memristor resistance. That is why the effect of line resistance is negligible. However, as the array size increases, the cumulative effect of the line resistance becomes more evident. The sneak path current is 3.42 $\mu$A for a 64×64 array with M6 interconnect compared to 1.44 $\mu$A in a 4×4 array, as shown in Fig.~\ref {fig: Fig8} (a). Although line resistance introduces measurable losses, it does not fundamentally alter the primary conclusion that sneak paths pose a significant challenge to the operation of large-scale memristor crossbar arrays. Similarly, the effect of selected line resistances (M3, M5, M6) for noise margin calculation is also shown in Fig.~\ref {fig: Fig8} (b). The line resistances are considerably smaller compared to the memristor resistances and load resistor used for noise margin calculation. It is evident in the figure that as the sneak path current increases in a large array, the noise margin drops accordingly. However, the noise margin values for different line resistances are similar due to the small numerical values of line resistances compared to higher memristor resistances and load resistance.

%\textcolor{red}{Fig. 5 is too small}

\begin{figure}
    \centering
    \includegraphics[width=1\linewidth]{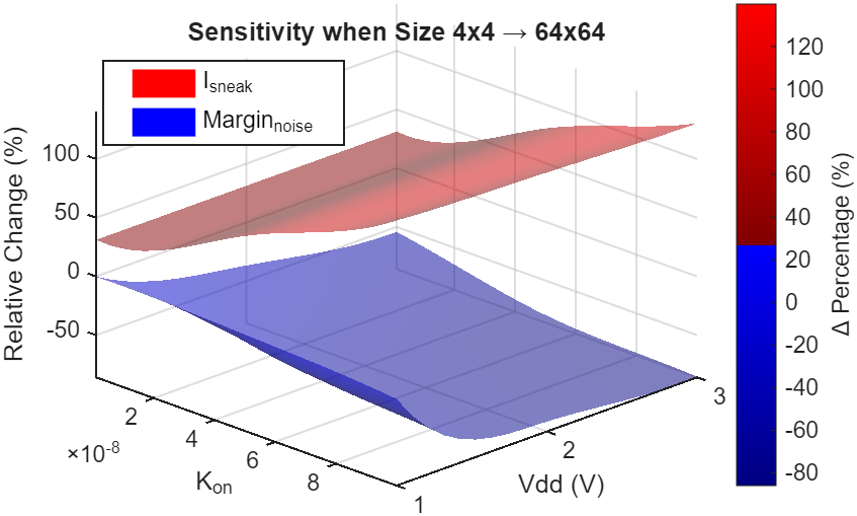}
    \caption{Relative percentage change in sneak path current and noise margin as K\textsubscript{on} and V\textsubscript{dd} are varied. The array adopts `All Ones' data pattern with FRC connection strategy using the 1.4nm technology node with Cu M3 interconnect.}
    \label{fig: Fig14}
\end{figure}

\subsection{ Model Validation Against SPICE Simulation} 
 To assess the accuracy and applicability of the closed-form model of the sneak path current, three sets of samples are selected for each parameter, and the resulting errors are analyzed. %\textcolor{blue}{The critical design parameters include the array size, data pattern, terminal connection strategy, ON/OFF ratio, interconnect resistance, and read voltage.} 
% \textcolor{blue}{The stray capacitance is 0.1fF.}
 For validation, intermediate points are selected instead of boundary points to avoid boundary effects, as fitting constraints at the boundaries may produce artificially small or near-zero errors. Table~\ref {tab:results2} presents the %\textcolor{blue}{
 SPICE
 %} 
 simulation results and model errors with high efficiency. All %\textcolor{blue}{
 errors
 %} 
 are %\textcolor{blue}{
 below
 %}
 %than 
 10.9\%,  indicating that the proposed model accurately captures the sneak path current. %\textcolor{blue}{
 In addition, under the same hardware environment, the proposed model achieves a run-time 10 to 4784 times faster than that of SPICE simulations. The hardware specifications are as follows: CPU Intel Core Ultra 7 256V, base frequency 2.2 GHz, 1 socket, 8 cores, 8 logical processors, and 16 GB of memory.%}

\subsection{Sensitivity Analysis} A detailed and systematic analysis of sneak path currents within passive memristor crossbar arrays is presented, thereby expanding the current research landscape regarding parameter coverage and the depth of analytical modeling. The parameters selected within this study aim to effectively connect circuit-level characteristics with fabrication-level considerations, thus providing a practical framework that integrates theoretical models with physical implementation. Sneak path currents are vital in evaluating the feasibility of array designs. Excessive sneak currents can significantly constrain the maximum achievable size of the array, decrease the noise margin of target cells, and hurt data integrity during read operations. Additionally, the ON/OFF resistance ratio of the devices plays an important role in influencing the severity of sneak paths: a lower contrast between the conductive and resistive states can mitigate undesired currents, whereas a higher ratio may intensify these currents. Moreover, this analysis explicitly considers the impact of interconnect line resistance, which becomes increasingly significant as devices and arrays continue to scale. In advanced technologies, line resistance contributes to voltage degradation along word lines and bit lines, thereby influencing write reliability and read stability. From the closed-form analysis, the observed errors result primarily from the limited number of sampled points in SPICE simulations. To balance run-time and model accuracy, the current number of sampling points is limited and array size is set from 4×4 to 64×64. Actually, the resulting errors could be further reduced by increasing the number of points or optimizing the model. Also, the close-form model has good scalability to be applicable for larger array size with the same methodology, which represents a promising direction for future work. Compared with full circuit-level simulations, the proposed model achieves comparable accuracy with substantially reduced time. The fast computing characteristics make the model suitable for large-scale array analysis and design optimization. To quantify the relative impact of array size on the noise margin and sneak path current, respectively, the size-to-margin sensitivity (Z\textsubscript{n}) and size-to-current sensitivity (Z\textsubscript{i}) for noise margin and sneak path current are defined in the following equations. A larger absolute value indicates a stronger sensitivity. 

\begin{equation}
%\begin{align}
    Z_n = \frac{ Margin_{noise\_64\times 64} - Margin_{noise\_4 \times 4}} { Margin_{noise\_4\times4}}
\end{equation}
%\end{align}

\begin{equation}
    Z_i = \frac{ I_{sneak\_64 \times 64} - I_{sneak\_4\times4}} { I_{sneak\_4\times4}} 
\end{equation}

\begin{figure}
    \centering
    \includegraphics[width=0.6\linewidth]{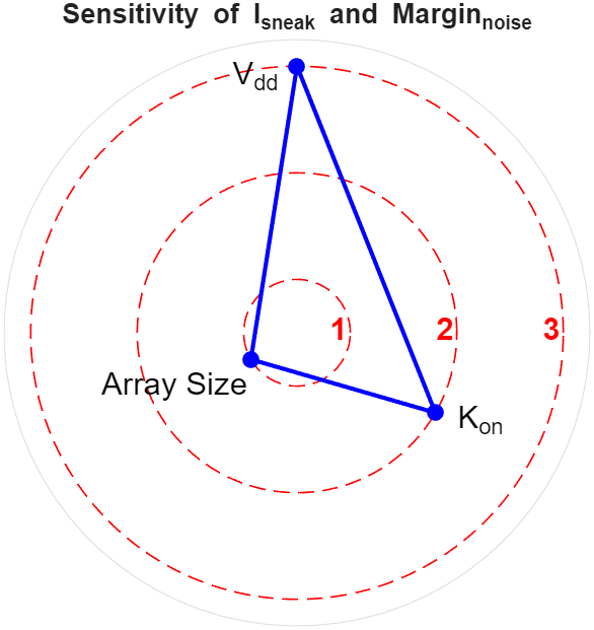}
    \caption{Radar chart for sensitivity of sneak path current and noise margin to key design parameters.} %in a memristor crossbar array.}
    \label{fig: Fig15}
\end{figure}

Taking M3 as an example, Fig.~\ref {fig: Fig14} presents the simulation results to quantify the tradeoff among critical design parameters on I\textsubscript{sneak} and noise margin.
The relative percentage change surface for both sneak current and noise margin is shown with respect to the change of V\textsubscript{dd} and K\textsubscript{on} values. In this work, the key input parameters across all connection strategies and data patterns are varied systematically, adjusting V\textsubscript{dd} by a factor of (3V - 1V)/1V = 2, K\textsubscript{on} for (1e-7-1e-9)/1e-9 = 99, and array size for (64×64 - 4×4)/(4×4) = 255. The resulting impact on sneak path current and noise margin is quantitatively ranked to assess parameter sensitivity. This analysis reveals how each parameter contributes to either aggravating or mitigating sneak path effects. Among the three, V\textsubscript{dd} exhibits the highest sensitivity, while array size shows the lowest. Such pre-design evaluation enables early prediction of how expanding the conductance range or lowering the read voltage can support larger array sizes, and vice versa. The trend and comparison for sensitivity are shown in a radar chart, Fig.~\ref {fig: Fig15}.%and V\textsubscript{dd},  K\textsubscript{on}, and array size are located in a radar map to show sensitivity comparison.

\section{Comparison with State-of-The-Art}
\label{sec:Comparison with State-of-The-Art}

\label{sec:comparison}

\begin{table*}[ht]
    \centering
    \caption{Comparison with State-of-the-Art}
    \label{table:comparison}
    \resizebox{0.6\linewidth}{!}{% <------ Don't forget this %
    \begin{tabular}{|c|c|c|c|c|c|}
    \hline
    Device Type & \cite{zidan2013memristor} & \cite{manem2010design} & \cite{joshi2021sneak} & \cite{datta2022unlocking} & Current \\
    \hline
    Array size consideration& $\surd$ & $\surd$ & $\surd$ & $\surd$ & $\surd$ \\ 
    \hline
    Read voltage consideration& $\surd$ & × & × & $\surd$ & $\surd$  \\ 
    \hline
    ON/OFF ratio consideration & $\surd$ & × & $\surd$ & $\surd$ & $\surd$ \\
    \hline
    Data pattern consideration& $\surd$  & × & $\surd$ & × & $\surd$\\
    \hline
    Connection strategy consideration& $\surd$ & $\surd$  & × & × & $\surd$\\
    \hline 
    Interconnect model consideration& $\surd$ & × & × & × & $\surd$\\
    \hline 
    Closed-form equation consideration& × & × & $\surd$ &× & $\surd$ \\
    \hline
    Accuracy consideration & × & × & × & × & $\surd$  \\
    \hline
    \end{tabular}%
    }
\end{table*}
 
Table~\ref{table:comparison}  provides a thorough comparison between the proposed analytical framework and existing studies on memristor-based systems. While previous research has addressed specific design parameters that influence sneak path behavior, none have comprehensively examined the complete set of interrelated factors affecting array performance. Key parameters, including array size, connection strategy, ON/OFF ratio, interconnect resistance, and read voltage, are essential for accurately evaluating the reliability and scalability of passive memristive memory arrays. In contrast, this study presents a novel and effective analytical approach that integrates all these factors within a cohesive framework. The proposed closed-form models demonstrate minimal error across various validation conditions, thus facilitating precise pre-design and real-time estimations, as well as enhanced performance predictions for memristor-based crossbar systems.

\section{Conclusion}
\label{sec:conclusion}

Sneak path currents significantly disrupt the reliable operation of passive memristor crossbar arrays. Therefore, estimating the sneak path effect during both the pre-design and ongoing design phases is essential for determining feasible array sizes and performance limitations. This work introduces an analytical framework for estimating sneak paths across a diverse range of design parameters and array configurations. The resulting closed-form models demonstrate the relationships of sneak path current with various parameters, revealing linear, exponential, and logarithmic dependencies under different biasing conditions. The proposed model is highly efficient, achieving an error of less than 10.9 \%\ while reducing computation time by a factor of 10 to 4784. This enables accurate and efficient estimation of the impact of sneak paths. The accompanying sensitivity analysis further highlights how variations in design parameters affect both sneak path current and noise margin, providing essential guidance for balancing competing design trade-offs. Such detailed assessments offer designers valuable insights into scalability, functional yield, and performance constraints for future memristor-based analog IMC systems. 

\section*{Acknowledgment}

This work is supported in part by the National Science Foundation under grants 2510192, 2502054, 2428981, 2420994, 2427766, 2408064, 2247343, and 2218046.  This material was also based upon work supported by the U.S. Department of Energy, Office of Science, Office of Advanced Scientific Computing Research (ASCR), under Award Number(s) DE-SC0025561.

%This work was supported in part by the National Science Foundation under Grants 2510192, 2502054, 2428981, 2420994, 2427766, 2408064, 2247343, and 2218046, and in part by the U.S. Department of Energy under Award DE-SC0025561.

\bibliographystyle{IEEEtran}

\bibliography{reference.bib}

@article{hu2018memristor,
  title={Memristor-based analog computation and neural network classification with a dot product engine},
  author={Hu, Miao and Graves, Catherine E and Li, Can and Li, Yunning and Ge, Ning and Montgomery, Eric and Davila, Noraica and Jiang, Hao and Williams, R Stanley and Yang, J Joshua and others},
  journal={Advanced Materials},
  volume={30},
  number={9},
  pages={1705914},
  year={2018},
  publisher={Wiley Online Library}
}

@inproceedings{andrulis2024cimloop,
  title={CiMLoop: A flexible, accurate, and fast compute-in-memory modeling tool},
  author={Andrulis, Tanner and Emer, Joel S and Sze, Vivienne},
  booktitle={2024 IEEE International Symposium on Performance Analysis of Systems and Software (ISPASS)},
  pages={10--23},
  year={2024},
  organization={IEEE}
}

@article{robinett2007demultiplexers,
  title={Demultiplexers for nanoelectronics constructed from nonlinear tunneling resistors},
  author={Robinett, Warren and Snider, Greg S and Stewart, Duncan R and Straznicky, Joseph and Williams, R Stanley},
  journal={IEEE transactions on nanotechnology},
  volume={6},
  number={3},
  pages={280--290},
  year={2007},
  publisher={IEEE}
}

@article{zhu2023mnsim,
  title={Mnsim 2.0: A behavior-level modeling tool for processing-in-memory architectures},
  author={Zhu, Zhenhua and Sun, Hanbo and Xie, Tongxin and Zhu, Yu and Dai, Guohao and Xia, Lixue and Niu, Dimin and Chen, Xiaoming and Hu, Xiaobo Sharon and Cao, Yu and others},
  journal={IEEE Transactions on Computer-Aided Design of Integrated Circuits and Systems},
  volume={42},
  number={11},
  pages={4112--4125},
  year={2023},
  publisher={IEEE}
}

@inproceedings{luo2019mlp+,
  title={MLP+ NeuroSimV3. 0: Improving on-chip learning performance with device to algorithm optimizations},
  author={Luo, Yandong and Peng, Xiaochen and Yu, Shimeng},
  booktitle={Proceedings of the international conference on neuromorphic systems},
  pages={1--7},
  year={2019}
}

@article{maram2023deep,
  title={A deep neural network deployment based on resistive memory accelerator simulation},
  author={Maram, Tejaswanth Reddy and Barnwal, Ria and others},
  journal={arXiv preprint arXiv:2304.11337},
  year={2023}
}

@inproceedings{manem2010design,
  title={Design considerations for variation tolerant multilevel CMOS/Nano memristor memory},
  author={Manem, Harika and Rose, Garrett S and He, Xiaoli and Wang, Wei},
  booktitle={Proceedings of the 20th symposium on Great lakes symposium on VLSI},
  pages={287--292},
  year={2010}
}

@article{joshi2021sneak,
  title={Sneak path characterization in memristor crossbar circuits},
  author={Joshi, Rasika and Acken, John M},
  journal={International Journal of Electronics},
  volume={108},
  number={8},
  pages={1255--1272},
  year={2021},
  publisher={Taylor \& Francis}
}

@inproceedings{datta2022unlocking,
  title={Unlocking sneak path analysis in memristor based logic design styles},
  author={Datta, Kamalika and Shirinzadeh, Saeideh and Thangkhiew, Phrangboklang Lyngton and Sengupta, Indranil and Drechsler, Rolf},
  booktitle={2022 25th Euromicro Conference on Digital System Design (DSD)},
  pages={793--800},
  year={2022},
  organization={IEEE}
}

@article{chua2003memristor,
  title={Memristor-the missing circuit element},
  author={Chua, Leon},
  journal={IEEE Transactions on circuit theory},
  volume={18},
  number={5},
  pages={507--519},
  year={2003},
  publisher={IEEE}
}

@inproceedings{uppaluru2025carbon,
  title={Carbon Efficiency of Natural Organic Honey-Memristor Based Neuromorphic Computing},
  author={Uppaluru, Harshvardhan and Templin, Zoe and Riam, Shah Zayed and Zhao, Feng and Wang, Jinhui},
  booktitle={Proceedings of the Great Lakes Symposium on VLSI 2025},
  pages={245--251},
  year={2025}
}

@incollection{vourkas2015memristive,
  title={Memristive crossbar-based nonvolatile memory},
  author={Vourkas, Ioannis and Sirakoulis, Georgios Ch},
  booktitle={Memristor-Based Nanoelectronic Computing Circuits and Architectures},
  pages={101--147},
  year={2015},
  publisher={Springer}
}

@article{shin2011analysis,
  title={Analysis of passive memristive devices array: Data-dependent statistical model and self-adaptable sense resistance for RRAMs},
  author={Shin, Sangho and Kim, Kyungmin and Kang, Sung-Mo},
  journal={Proceedings of the IEEE},
  volume={100},
  number={6},
  pages={2021--2032},
  year={2011},
  publisher={IEEE}
}

@article{kim2012functional,
  title={A functional hybrid memristor crossbar-array/CMOS system for data storage and neuromorphic applications},
  author={Kim, Kuk-Hwan and Gaba, Siddharth and Wheeler, Dana and Cruz-Albrecht, Jose M and Hussain, Tahir and Srinivasa, Narayan and Lu, Wei},
  journal={Nano letters},
  volume={12},
  number={1},
  pages={389--395},
  year={2012},
  publisher={ACS Publications}
}

@article{jo2009high,
  title={High-density crossbar arrays based on a Si memristive system},
  author={Jo, Sung Hyun and Kim, Kuk-Hwan and Lu, Wei},
  journal={Nano letters},
  volume={9},
  number={2},
  pages={870--874},
  year={2009},
  publisher={ACS Publications}
}

@article{lastras2018resistive,
  title={Resistive random-access memory based on ratioed memristors},
  author={Lastras-Montano, Miguel Angel and Cheng, Kwang-Ting},
  journal={Nature Electronics},
  volume={1},
  number={8},
  pages={466--472},
  year={2018},
  publisher={Nature Publishing Group UK London}
}

@article{vontobel2009writing,
  title={Writing to and reading from a nano-scale crossbar memory based on memristors},
  author={Vontobel, Pascal O and Robinett, Warren and Kuekes, Philip J and Stewart, Duncan R and Straznicky, Joseph and Williams, R Stanley},
  journal={Nanotechnology},
  volume={20},
  number={42},
  pages={425204},
  year={2009},
  publisher={IOP Publishing}
}

@inproceedings{cassuto2013sneak,
  title={Sneak-path constraints in memristor crossbar arrays},
  author={Cassuto, Yuval and Kvatinsky, Shahar and Yaakobi, Eitan},
  booktitle={2013 IEEE international symposium on information theory},
  pages={156--160},
  year={2013},
}

@article{gul2019addressing,
  title={Addressing the sneak-path problem in crossbar RRAM devices using memristor-based one Schottky diode-one resistor array},
  author={G{\"u}l, Fatih},
  journal={Results in Physics},
  volume={12},
  pages={1091--1096},
  year={2019},
  publisher={Elsevier}
}

@article{tang2018comprehensive,
  title={Comprehensive sensing current analysis and its guideline for the worst-case scenario of RRAM read operation},
  author={Tang, Zhensen and Wang, Yao and Chi, Yaqing and Fang, Liang},
  journal={Electronics},
  volume={7},
  number={10},
  pages={224},
  year={2018},
  publisher={MDPI}
}

@inproceedings{liang2010size,
  title={Size limitation of cross-point memory array and its dependence on data storage pattern and device parameters},
  author={Liang, Jiale and Wong, H-S Philip},
  booktitle={2010 IEEE International Interconnect Technology Conference},
  pages={1--3},
  year={2010},
}

@article{pei2024ultra,
  title={Ultra-Scaled E-Tree-Based SRAM Design and Optimization With Interconnect Focus},
  author={Pei, Zhenlin and Liu, Hsiao-Hsuan and Mayahinia, Mahta and Tahoori, Mehdi B and Catthoor, Francky and T{\H{o}}kei, Zsolt and Abdi, Dawit Burusie and Myers, James and Pan, Chenyun},
  journal={IEEE Transactions on Circuits and Systems I: Regular Papers},
  year={2024},
  publisher={IEEE}
}

@article{pei2025interconnect,
  title={Interconnect/memory co-design and co-optimization using differential transmission lines},
  author={Pei, Zhenlin and Liu, Hsiao-Hsuan and Mayahinia, Mahta and Tahoori, Mehdi and Catthoor, Francky and Tokei, Zsolt and Dubey, Prashant and Pan, Chenyun},
  journal={IEEE Transactions on Very Large Scale Integration (VLSI) Systems},
  year={2025},
  publisher={IEEE}
}

@article{liu2023cfet,
  title={CFET SRAM DTCO, interconnect guideline, and benchmark for CMOS scaling},
  author={Liu, Hsiao-Hsuan and Salahuddin, Shairfe M and Chan, Boon Teik and Schuddinck, Pieter and Xiang, Yang and Hellings, Geert and Weckx, Pieter and Ryckaert, Julien and Catthoor, Francky},
  journal={IEEE Transactions on Electron Devices},
  volume={70},
  number={3},
  pages={883--890},
  year={2023},
  publisher={IEEE}
}

@inproceedings{pei2023technology,
  title={Technology/memory co-design and co-optimization using E-tree interconnect},
  author={Pei, Zhenlin and Mayahinia, Mahta and Liu, Hsiao-Hsuan and Tahoori, Mehdi and Catthoor, Francky and Tokei, Zsolt and Pan, Chenyun},
  booktitle={Proceedings of the Great Lakes Symposium on VLSI 2023},
  pages={159--162},
  year={2023}
}

@article{liu2024future,
  title={Future design direction for SRAM data array: Hierarchical subarray with active interconnect},
  author={Liu, Hsiao-Hsuan and Gilardi, Carlo and Salahuddin, Shairfe M and Pei, Zhenlin and Schuddinck, Pieter and Xiang, Yang and Weckx, Pieter and Hellings, Geert and Bardon, Marie Garcia and Ryckaert, Julien and others},
  journal={IEEE Transactions on Circuits and Systems I: Regular Papers},
  volume={71},
  number={12},
  pages={6495--6506},
  year={2024},
  publisher={IEEE}
}

@article{liu2023cfetted,
  title={CFET SRAM with double-sided interconnect design and DTCO benchmark},
  author={Liu, Hsiao-Hsuan and Schuddinck, Pieter and Pei, Zhenlin and Verschueren, Lynn and Mertens, Hans and Salahuddin, Shairfe M and Hiblot, Gaspard and Xiang, Yang and Chan, Boon Teik and Subramanian, Sujith and others},
  journal={IEEE Transactions on Electron Devices},
  volume={70},
  number={10},
  pages={5099--5106},
  year={2023},
  publisher={IEEE}
}

@article{chen2013comprehensive,
  title={A comprehensive crossbar array model with solutions for line resistance and nonlinear device characteristics},
  author={Chen, An},
  journal={IEEE Transactions on Electron Devices},
  volume={60},
  number={4},
  pages={1318--1326},
  year={2013},
  publisher={IEEE}
}

@article{zidan2013memristor,
  title={Memristor-based memory: The sneak paths problem and solutions},
  author={Zidan, Mohammed Affan and Fahmy, Hossam Aly Hassan and Hussain, Muhammad Mustafa and Salama, Khaled Nabil},
  journal={Microelectronics journal},
  volume={44},
  number={2},
  pages={176--183},
  year={2013},
  publisher={Elsevier}
}

@article{linn2010complementary,
  title={Complementary resistive switches for passive nanocrossbar memories},
  author={Linn, Eike and Rosezin, Roland and K{\"u}geler, Carsten and Waser, Rainer},
  journal={Nature materials},
  volume={9},
  number={5},
  pages={403--406},
  year={2010},
  publisher={Nature Publishing Group UK London}
}

@article{shi2020research,
  title={Research progress on solutions to the sneak path issue in memristor crossbar arrays},
  author={Shi, Lingyun and Zheng, Guohao and Tian, Bobo and Dkhil, Brahim and Duan, Chungang},
  journal={Nanoscale Advances},
  volume={2},
  number={5},
  pages={1811--1827},
  year={2020},
  publisher={Royal Society of Chemistry}
}

@inproceedings{kannan2013sneak,
  title={Sneak-path testing of memristor-based memories},
  author={Kannan, Sachhidh and Rajendran, Jeyavijayan and Karri, Ramesh and Sinanoglu, Ozgur},
  booktitle={2013 26th International Conference on VLSI Design and 2013 12th International Conference on Embedded Systems},
  pages={386--391},
  year={2013},
}

@article{strukov2008missing,
  title={The missing memristor found},
  author={Strukov, Dmitri B and Snider, Gregory S and Stewart, Duncan R and Williams, R Stanley},
  journal={nature},
  volume={453},
  number={7191},
  pages={80--83},
  year={2008},
  publisher={Nature Publishing Group UK London}
}

@article{demin2020sneak,
  title={Sneak, discharge, and leakage current issues in a high-dimensional 1T1M memristive crossbar},
  author={Demin, VA and Surazhevsky, IA and Emelyanov, AV and Kashkarov, PK and Kovalchuk, MV},
  journal={Journal of Computational Electronics},
  volume={19},
  number={2},
  pages={565--575},
  year={2020},
  publisher={Springer}
}

@inproceedings{pourbakhsh2016sizing,
  title={Sizing-priority based low-power embedded memory for mobile video applications},
  author={Pourbakhsh, Seyed Alireza and Chen, Xiaowei and Chen, Dongliang and Wang, Xin and Gong, Na and Wang, Jinhui},
  booktitle={2016 17th International Symposium on Quality Electronic Design (ISQED)},
  pages={1--5},
  year={2016},
}

@article{gong2012clock,
  title={Clock-biased local bit line for high performance register files},
  author={Gong, Na and Wang, Jinhui and Jiang, Shixiong and Sridhar, R},
  journal={Electronics letters},
  volume={48},
  number={18},
  pages={1104--1105},
  year={2012},
  publisher={IET}
}

@ARTICLE{6730965,
  author={Gong, Na and Wang, Jinhui and Sridhar, Ramalingam},
  journal={IEEE Transactions on Circuits and Systems I: Regular Papers}, 
  title={Variation Aware Sleep Vector Selection in Dual ${\rm V}_{{{\rm t}}}$ Dynamic OR Circuits for Low Leakage Register File Design}, 
  year={2014},
  volume={61},
  number={7},
  pages={1970-1983}
}

@article{oli2022stuck,
  title={Stuck-at-fault immunity enhancement of memristor-based edge ai systems},
  author={Oli-Uz-Zaman, Md and Khan, Saleh Ahmad and Oswald, William and Liao, Zhiheng and Wang, Jinhui},
  journal={IEEE Journal on Emerging and Selected Topics in Circuits and Systems},
  volume={12},
  number={4},
  pages={922--933},
  year={2022},
  publisher={IEEE}
}

@inproceedings{oli2022reliability,
  title={Reliability improvement in rram-based dnn for edge computing},
  author={Oli-Uz-Zaman, Md and Khan, Saleh Ahmad and Yuan, Geng and Wang, Yanzhi and Liao, Zhiheng and Fu, Jingyan and Ding, Caiwen and Wang, Jinhui},
  booktitle={2022 IEEE international symposium on circuits and systems (ISCAS)},
  pages={581--585},
  year={2022},
}

@article{edstrom2019data,
  title={Data-Pattern Enabled Self-Recovery Low-Power Storage System for Big Video Data},
  author={Edstrom, Jonathon and Chen, Dongliang and Gong, Yifu and Wang, Jinhui and Gong, Na},
  journal={IEEE Transactions on Big Data},
  volume={5},
  number={01},
  pages={95--105},
  year={2019},
  publisher={IEEE Computer Society}
}

@article{gong2014variation,
  title={Variation aware sleep vector selection in dual Vt dynamic or circuits for low leakage register file design},
  author={Gong, Na and Wang, Jinhui and Sridhar, Ramalingam},
  journal={IEEE Transactions on Circuits and Systems I: Regular Papers},
  volume={61},
  number={7},
  pages={1970--1983},
  year={2014},
  publisher={IEEE}
}

@article{chen2018viewer,
  title={Viewer-aware intelligent efficient mobile video embedded memory},
  author={Chen, Dongliang and Edstrom, Jonathon and Gong, Yifu and Gao, Peng and Yang, Lei and McCourt, Mark E and Wang, Jinhui and Gong, Na},
  journal={IEEE Transactions on Very Large Scale Integration (VLSI) Systems},
  volume={26},
  number={4},
  pages={684--696},
  year={2018},
  publisher={IEEE}
}

@article{khan2023pawn,
  title={Pawn: Programmed analog weights for non-linearity optimization in memristor-based neuromorphic computing system},
  author={Khan, Saleh Ahmad and Oli-Uz-Zaman, Md and Wang, Jinhui},
  journal={IEEE Journal on Emerging and Selected Topics in Circuits and Systems},
  volume={13},
  number={1},
  pages={436--444},
  year={2023},
  publisher={IEEE}
}

\vspace{-0.3in}

\begin{IEEEbiography}[{\includegraphics[width=1\linewidth,height=1.3\linewidth, clip,]{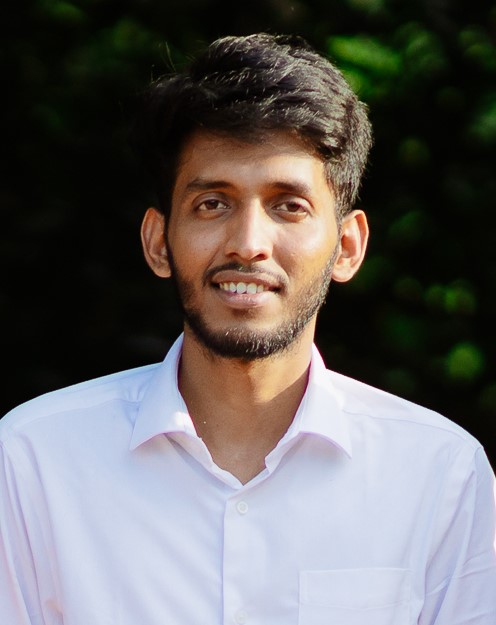}}]{Shah Zayed Riam}   received the B.Sc. and M.S. degrees in electrical engineering from Rajshahi University of Engineering and Technology (RUET), Bangladesh, and the University of Texas at Tyler, USA, respectively, in 2022 and 2024. He is currently pursuing a Ph.D. degree at the University of Alabama, Tuscaloosa, AL, USA. His research interests include memory design, flexible electronics, and electrochemical biosensors. His previous works received a runner-up award at WiSe/YP BIP 2023 and a nomination for the Best Paper Award at IEEE FLEPS 2023, Boston, MA, USA.
\end{IEEEbiography}

\vspace{-.3in}

\begin{IEEEbiography}[{\includegraphics[width=1\linewidth,height=1.25\linewidth,clip]{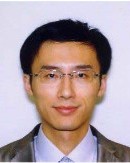}}]{Zhenlin Pei} (Graduate Student Member, IEEE) holds the M.S. and Ph.D. degrees in electrical engineering from Columbia University and the University of Texas at Arlington. During his Ph.D., he collaborated with the Interuniversity Microelectronics Centre to develop the Cacti++ framework, which bridges gaps in rapid EDA-based co-design and co-optimization from the transistor to the system level. He is a postdoctoral fellow in the department of electrical \& computer engineering at the University of Alabama. He was a senior design engineer in the IP group for tapeout at Cadence for four years. Research interests include energy-efficient computing from hardware to software, with applications in AI through CAD/EDA \& DTCO/STCO, emerging interconnects/beyond-CMOS/memory technologies, AI hardware with privacy, and neuromorphic systems.
\end{IEEEbiography}

\vspace{-.3in}

\begin{IEEEbiography}[{\includegraphics[width=1\linewidth,height=1.25\linewidth,clip]{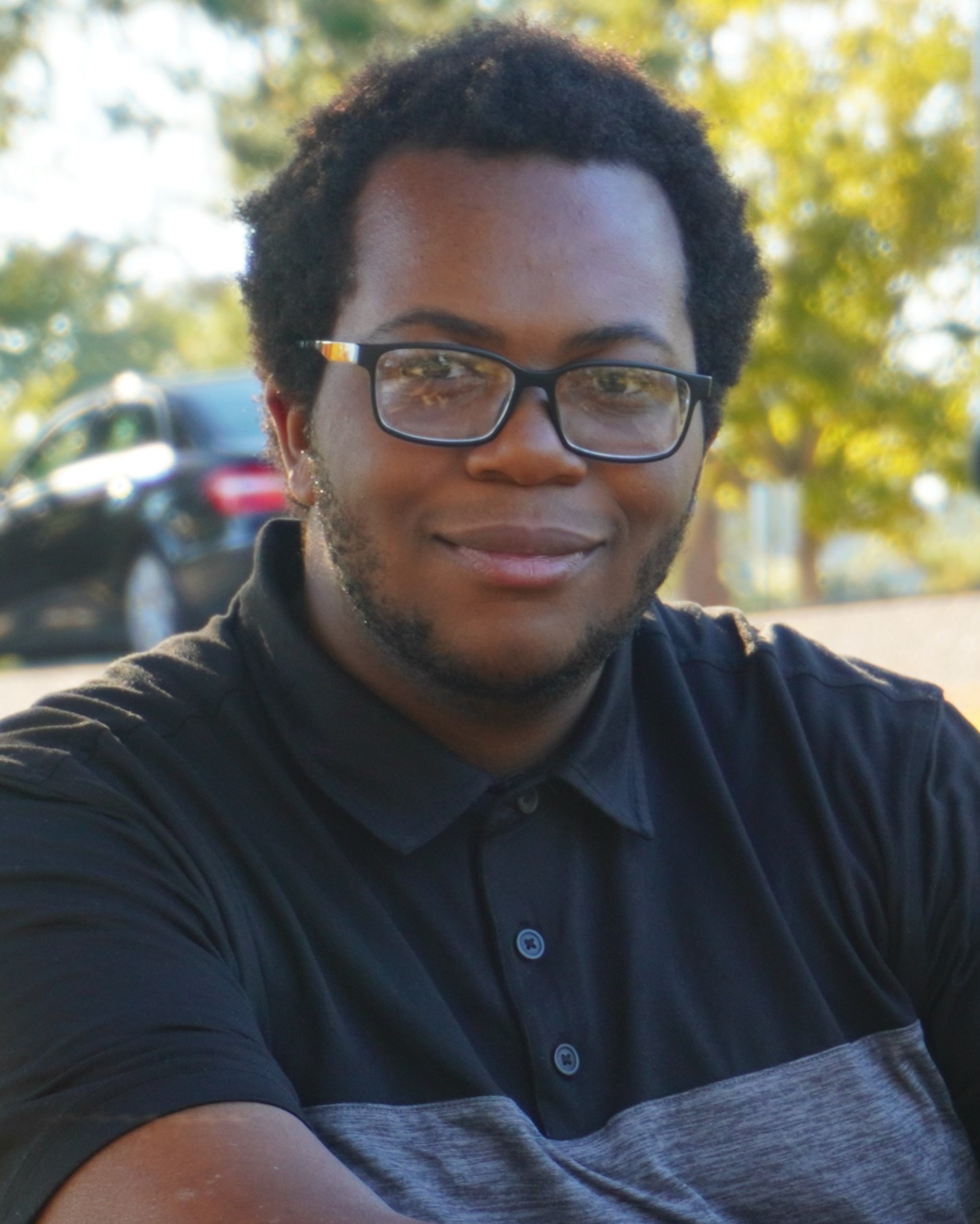}}]{Kyle Mooney} (Graduate Student Member, IEEE) received his B.S. degree in computer engineering from the University of South Alabama in 2023 and is currently pursuing a Ph.D. in Electrical Engineering from the University of Alabama. His research interests include memory design, edge computing, and artificial intelligence.
\end{IEEEbiography}

\vspace{-.3in}

\begin{IEEEbiography}[{\includegraphics[width=1\linewidth,height=1.25\linewidth,clip]{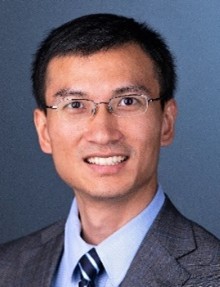}}]{Chenyun Pan}  (Senior Member, IEEE) received a B.S. in Microelectronics from Shanghai Jiao Tong University and a Ph.D. in ECE from Georgia Institute of Technology. He is an Associate Professor at the University of Texas at Arlington, and his research focuses on energy-efficient Boolean and non-Boolean computing systems using emerging technologies. He has published over 80 IEEE papers and received several awards, including the Research Spotlight Award from Georgia Institute of Technology, and an early career research award from the U.S. Department of Energy.
\end{IEEEbiography}

\vspace{-.3in}

\begin{IEEEbiography}[{\includegraphics[width=1\linewidth,height=1.25\linewidth,clip]{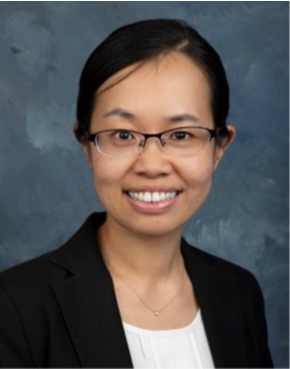}}]{Na Gong} (Senior Member, IEEE) received the Ph.D. degree in computer science and engineering from the State University of New York, Buffalo, in 2013. Currently, Dr. Gong is a professor in the Department of Electrical and Computer Engineering at the University of Alabama. Her research interests include power-efficient computing circuits and systems, memory optimization, AI hardware, and hardware privacy. She is the recipient of the best paper nomination from ISVLSI’19, best paper award from EIT’16, best paper nominations from ISQED’16 and ISLPED’16.
\end{IEEEbiography}

\vspace{-.3in}

\begin{IEEEbiography}[{\includegraphics[width=1\linewidth,height=1.25\linewidth,clip]{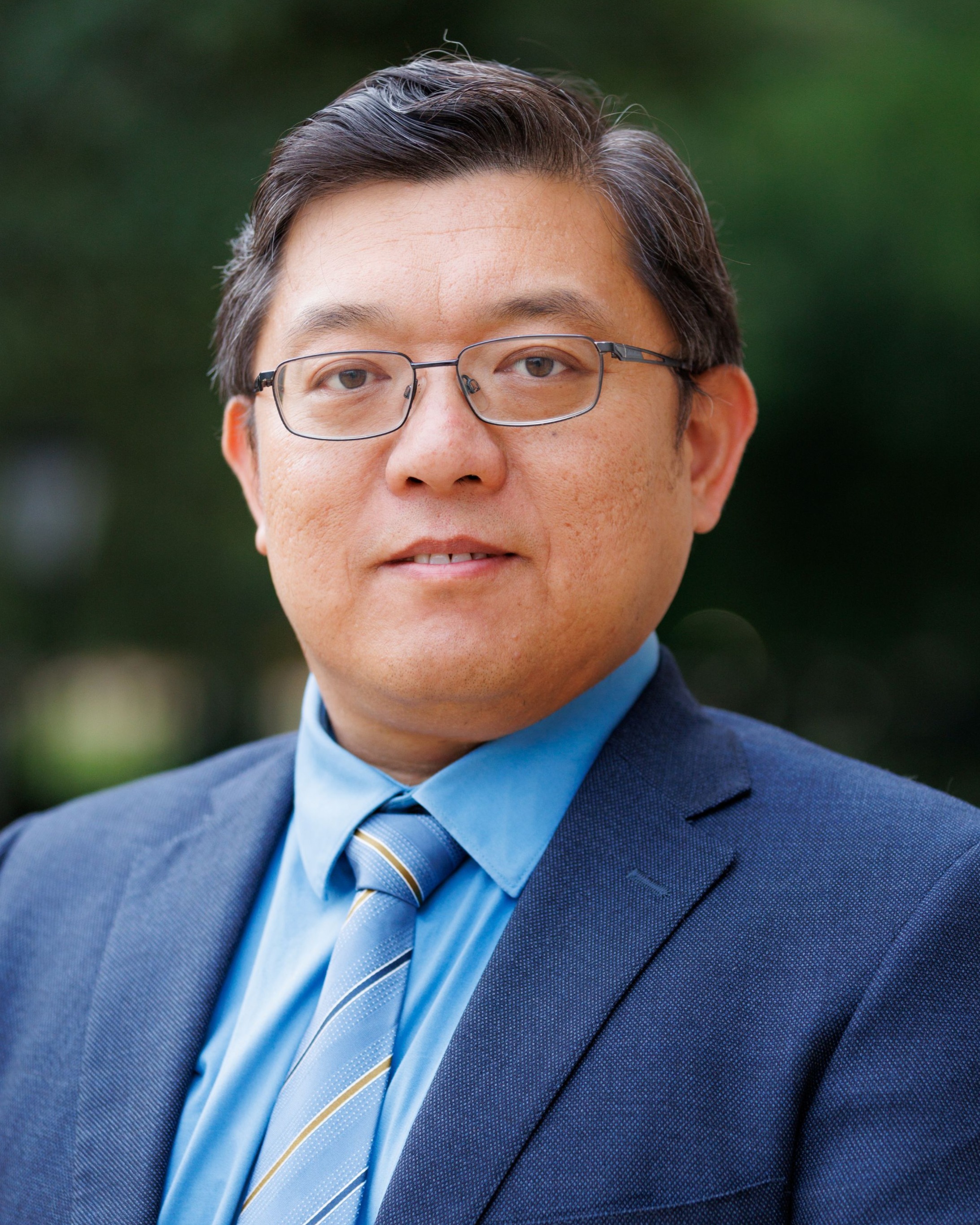}}]{Jinhui Wang} (Senior Member, IEEE) is currently a Full Professor and Larry Drummond Endowed Chair with the Department of Electrical and Computer Engineering at the University of Alabama, Tuscaloosa, AL, USA. His research interests include: (1) VLSI System, Digital and Mixed-Signal Integrated Circuit (IC) Design, 3D and 2.5D IC Design, and Emerging Memory; (2) AI Hardware Design, Post/Beyond CMOS Device, such as Memristors, Based Neuromorphic Computing System; and (3) Post/Beyond CMOS Devices Enabled Cybersecurity and Internet of Things (IoT) Systems. He has published over 200 refereed journal/conference papers and book chapters as well as 31 patents in the area of emerging semiconductor technologies. His previous work has received the Best Paper Award/Nomination at DATE 2021, ISVLSI 2019, ISLPED 2016, ISQED 2016, and EIT 2016.
\end{IEEEbiography}

\vfill

\end{document}